\documentclass[floatfix,reprint,amsmath,amssymb,aps,prb,superscriptaddress]{revtex4-2}

\usepackage{preamble}

\begin{document}

\title{Vortex NOON states for rotation sensing}

\author{Simon Dengis}
\email{simon.dengis@lkb.upmc.fr}
\affiliation{CESAM research unit, University of Li\`ege, B-4000 Li\`ege, Belgium}
\affiliation{Laboratoire Kastler Brossel, Sorbonne Université, CNRS, ENS-Universit\'e PSL, Collège de France, 4 Place Jussieu, 75005 Paris, France}

\author{Nathan Dupont}
\email{nathan.dupont@lkb.upmc.fr}
\affiliation{Laboratoire Kastler Brossel, Sorbonne Université, CNRS, ENS-Universit\'e PSL, Collège de France, 4 Place Jussieu, 75005 Paris, France}
\affiliation{Center for Nonlinear Phenomena and Complex Systems, Université Libre de Bruxelles, CP 231, Campus Plaine, 1050 Brussels, Belgium}
\affiliation{International Solvay Institutes, 1050 Brussels, Belgium}

\author{Peter Schlagheck}
\email{peter.schlagheck@uliege.be}
\affiliation{CESAM research unit, University of Li\`ege, B-4000 Li\`ege, Belgium}

\author{Nathan Goldman}
\email{nathan.goldman@lkb.ens.fr}
\thanks{\newline SD and ND contributed equally to this work.}
\affiliation{Center for Nonlinear Phenomena and Complex Systems, Université Libre de Bruxelles, CP 231, Campus Plaine, 1050 Brussels, Belgium}
\affiliation{International Solvay Institutes, 1050 Brussels, Belgium}
\affiliation{Laboratoire Kastler Brossel, Coll\`ege de France, CNRS, ENS-Universit\'e PSL, Sorbonne Universit\'e, 11 Place Marcelin Berthelot, 75005 Paris, France}

\begin{abstract}
We introduce a scheme to generate NOON states of few-body bosonic vortices and demonstrate their application as high-precision rotation sensors. Our approach is based on cold atoms in a weakly anisotropic two-dimensional harmonic trap, where the single-particle p orbitals define an effective two-mode Bose–Hubbard model with vortex modes $(\text{p}_x\pm\di\text{p}_y)$ carrying opposite circulation. In the self-trapping regime, we show that the NOON manifold becomes spectrally isolated, and collective tunneling processes give rise to highly entangled vortex NOON states. However, these states emerge on prohibitively long timescales. To overcome this limitation, we develop two complementary acceleration strategies: geodesic counterdiabatic driving for small particle numbers, and resonance- and chaos-assisted tunneling in the semiclassical regime at larger particle numbers. Both approaches enable the generation of NOON states on experimentally relevant timescales while preserving near-unit fidelities. Finally, we quantify the metrological advantage of vortex NOON states by introducing an interferometric protocol that exploits their intrinsic sensitivity to rotation, enabling the detection of infinitesimal external rotations at the Heisenberg limit. Our work opens the door to rotation sensors based on atomic NOON states, generically realizable in bosonic Josephson junctions with vortex‑type orbitals.
\end{abstract}
\maketitle

\section{Introduction}

\begin{figure}[!t]
    \centering
    \includegraphics[width=0.475\textwidth]{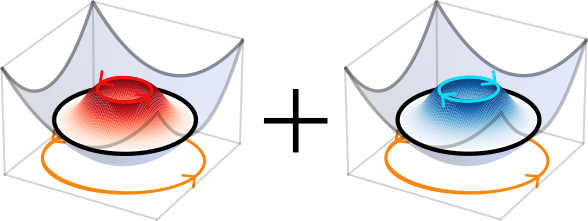}
    \caption{
    \textbf{Vortex NOON states for rotation sensing.}
    Schematic representation of an ensemble of bosons confined in a two-dimensional harmonic trap (gray parabolic area). The goal is to create within this system a NOON state of vortices, $\ket{\dNOON} = (\ket{N,0} + \ket{0,N})/\sqrt{2} $, a superposition of one component where all bosons circulate in one direction (red) and another where they all circulate in the opposite direction (blue). Here, $\ket{n_+,n_-}$ denotes the Fock state with $n_\pm$ bosons in the vortex mode $\ket{\psi_\pm}$~\eqref{eq:psi_pm}. In principle, the enhanced sensitivity of the NOON states could enable the measurement of a weak external rotation (orange arrows; e.g.~Earth's rotation) through a suitable metrology protocol.}
    \label{fig:schema}
\end{figure}

Quantum metrology aims at harnessing nonclassical features of quantum systems to estimate physical parameters more precisely than any classical strategy allows. In particular, quantum entanglement between the particles of a probe can sharpen this precision well beyond the standard quantum limit (the $1/\sqrt{N}$ scaling set by $N$ independent particles) down to the ultimate Heisenberg limit, scaling as $1/N$~\cite{degen_2017, pezze_2018, Pezze_2020}. Paradigmatic examples are NOON states (see Fig.~\ref{fig:schema}), coherent superpositions of all $N$ particles occupying one mode and all $N$ occupying the other, $(\ket{N,0}+\ket{0,N})/\sqrt{2}$, whose phase sensitivity saturates this Heisenberg limit~\cite{Kwon_2019, Jun_2024, Panda_2024}. This makes them appealing probes for interferometric measurements of quantities such as optical phase, frequency, and rotation \cite{pezze_2008, jones_2009, Hallwood2009, Humphreys_2013,Pezze_2014, Zhang_2018}.
For the latter, NOON states of vortices are particularly well-suited: since the two modes of the system carry opposite angular momenta (see Fig.~\ref{fig:schema}), an externally applied rotation at angular velocity $\Omega$ induces a relative phase shift between the two components. This accumulated phase is proportional both to the angular momentum and to the interrogation time, making such states promising candidates for ultra-precise rotation sensing \cite{Canuel_2006,Kumar_2021,roy_2026}.

However, the extreme susceptibility of NOON states to decoherence and losses makes their generation and control a major experimental challenge~\cite{li_2008,pezze_2018}. As a result, considerable effort has been devoted to the search for platforms and protocols capable of robustly producing high-fidelity NOON states on an experimentally feasible timescale~\cite{degen_2017,pezze_2018}. While NOON states have been recently realized experimentally with bosonic particles such as phonons or photons \cite{Itai_2010, Song_2017, Zhang_2018}, the realization of such entangled states with ultracold atoms is still the subject of extensive discussion \cite{Cirac_1998, Gordon_1999, Sorensen_2001, Mahmud_2003,Schenke2011, Fischer_2015, Bycheck_2018, Pezze_2019, Carr_2010, Schneider_2022, Beringer_2024}. In this context, the presence of interatomic interactions offers new perspectives for implementation protocols, as the dynamics of such a system differs greatly from that observed in photonic or phononic systems. When ultracold atoms are trapped in a lattice, atom-atom interactions can be exploited to define a subspace of accessible states that is isolated from the rest of the Hilbert space. Indeed, the energy levels in which all particles occupy the same mode can be separated from the rest of the spectrum when the system lies in the self-trapping regime \cite{smerzi_1997, Milburn_1997, albiez_2005, Carr_2010}, where interatomic interactions dominate over the tunneling amplitude of the bosons. In this regime, sequential tunneling is suppressed and the bosons may only undergo collective tunneling. This phenomenon enabled the first experimental realization of NOON-like states in ultracold atoms trapped in a symmetric double well \cite{Zhang_2025}, though achieving high-fidelity states remains challenging:~relying solely on collective tunneling does not yield sufficiently pure entangled states within experimentally accessible timescales. To overcome this limitation, several methods of quantum control exist, such as optimal control \cite{lapert_2012,Koch_2022,Beringer_2024}, shortcuts-to-adiabaticity \cite{Chen_2010, Chen_2010_b, delCampo_2013, Deffner_2014, delCampo_2019,Guery-Odelin_2019,Cepaite_2023} and resonance- and chaos-assisted tunneling \cite{tomsovic_1994,Brodier_2001, eltschka_2005,schlagheck_2011}. 

In this work, we propose a theoretical scheme for the creation and use of vortex NOON states in a weakly anisotropic two-dimensional harmonic trap. In this setting, the two lowest excited single-particle states (``p states'') define an effective two-mode Bose-Hubbard model, whose modes correspond to opposite circulation of the bosons \cite{li_2016}. In the self-trapping regime, interactions isolate the NOON manifold and collective tunneling naturally generates coherent superpositions of the two vortex modes. As we demonstrate in this work, these states constitute particularly appealing resources for rotation sensing, as their opposite angular momenta directly encode an external rotation into a measurable relative phase through a simple interferometric sequence.

A central challenge is that collective tunneling becomes exponentially slow with increasing particle number, making direct state preparation impractical. To accelerate the tunneling process, we investigate two complementary mechanisms that were applied to similar systems in the recent years \cite{vanhaele_2021, vanhaele_2022, Dengis_2025_a, Dengis_2025_b}. For small particle numbers $N\leq 8$, we show that geodesic counterdiabatic driving enables the deterministic preparation of high-fidelity vortex NOON states on experimentally relevant timescales \cite{Dengis_2025_a, Dengis_2025_b}. For larger particle numbers, we exploit resonance- and chaos-assisted tunneling induced by a periodic modulation of the rotation, leading to a strong enhancement of the tunneling dynamics in the semiclassical regime \cite{vanhaele_2021,vanhaele_2022}. Together, these two approaches provide a route toward the efficient generation of vortex entanglement and its application to quantum-enhanced rotation sensing. While this work focuses on the vortex states supported by a 2D trap, we emphasize that the NOON-state preparation and metrological schemes developed here apply equally well to other effective two‑mode settings exhibiting opposite circulating currents, such as ring‑shaped BECs threaded by a synthetic flux~\cite{aghamalyan2015coherent, Nicolau_2020,Pradhan_2024, Carmona-Lopez2026} and $\pi$-flux plaquettes~\cite{DiLiberto_pi_flux}.

The next sections are organized as follows. Section \ref{sec:bh_model} describes the theoretical framework of bosonic vortices trapped in an anisotropic 2D potential. An interferometry scheme for harnessing the high sensitivity of NOON states to rotation sensing is proposed. We then introduce the self-trapping regime and the mean-field limit of the model. Section \ref{sec:gcd} develops a creation scheme for $N\leq 8$ particles by means of the geodesic counterdiabatic driving, and we show how to implement this type of protocol experimentally. Section \ref{sec:RATCAT} is focused on the creation of NOON states for larger number of particles ($N\geq 8$) by enhancing collective tunneling using resonance- and chaos-assisted tunneling.

\section{Bosonic vortices in an anisotropic 2D trap}

\subsection{System and Bose-Hubbard model}
\label{sec:bh_model}

\begin{figure}
    \centering
    \includegraphics[scale=1]{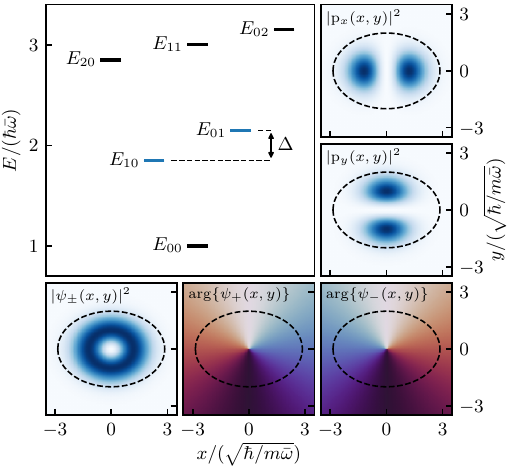}
    \caption{
    \textbf{One-body subspace.} Top left: First six levels $E_{n_x,n_y}$ of the 2D quantum harmonic oscillator Eq.~\eqref{eq:H0_1body} for $\Omega=0$ with anisotropy $\Delta/\hbar\bar{\omega}$ (exaggerated for clarity compared to the weakly anisotropic regime considered; see Eqs.~\eqref{eq:anisotropy} and~\eqref{eq:weak_anisotropy}) and subspace of interest $\{\ket{\dpx},\ket{\dpy}\}$ in blue. Five remaining panels: $|\dpx(x,y)|^2$, $|\dpy(x,y)|^2$ and their vortex superpositions $|\psi_\pm(x,y)|^2$ (see Eq.~\eqref{eq:psi_pm}) with their phase (curling in opposite direction). The dashed lines correspond to the iso-potential \mbox{$U(x,y) = 2 \hbar \bar{\omega}$}.
    }
    \label{fig:2}
\end{figure}

We consider an ensemble of $N$ interacting bosons in an anisotropic two-dimensional (2D) harmonic potential of angular frequencies $(\omega_x,\omega_y)$, which rotates at an angular frequency $\Omega$.
At the one-body level, the Hamiltonian describing a boson of mass $m$ in the frame co-rotating with the trap is~\cite{crepel_2023}:
\begin{equation}
    \label{eq:H0_1body}
    \hH_\ob = \dfrac{\hp_x^2+\hp_y^2}{2m} + \dfrac{m}{2} \left( \omega_x^2 \hx^2 + \omega_y^2 \hy^2 \right) - \Omega \hL_z^{(\ob)},
\end{equation}
where \mbox{$\hL_z^{(\ob)} = \hx\hp_y - \hy\hp_x$} is the angular momentum operator acting at the one-body level, and $\Omega < \min(\wx,\wy)$ such that the trap is confining.
Using the ladder operators $\ha_q$ and $\ha_q^\dagger$ (with \mbox{$\ha_q = \sqrt{m\omega_q/2\hbar} (\hat{q} + \di \hp_q / m \omega_q)$} and $q = x$ or $y$) this one-body Hamiltonian becomes
\begin{equation}
\label{eq:H0_1body_ladder}
\begin{split}
    \hH_\ob = \, &\hbar \wx\left(\hn_x + \dfrac{1}{2}\right) + \hbar \wy\left(\hn_y+ \dfrac{1}{2}\right)\\
    &- \di \hbar \Omega\dfrac{\wx+\wy}{2\sqrt{\wx\wy}}\left(\ha_y^\dagger \ha_x- \ha_x^\dagger\ha_y \right) \\
    &- \di \hbar \Omega\dfrac{\wy-\wx}{2\sqrt{\wx\wy}}\left(\ha_x^\dagger\ha_y^\dagger - \ha_x\ha_y\right).
\end{split}
\end{equation}
In the absence of external rotation ($\Omega = 0$), the eigenstates of this 2D harmonic oscillator (HO) are the states \mbox{$|\phi_{n_x n_y}\rangle \equiv|\phi_{n_x}\rangle|\phi_{n_y}\rangle$}, whose eigenenergies $E_{n_x,n_y}$ are given by the first line of Eq.~\eqref{eq:H0_1body_ladder}. For $\Omega\neq 0$, $\hL_z^{(\ob)}$ couples the states $|\phi_{n_x n_y}\rangle$. We consider the regime of weak anisotropy:
\begin{equation}
    \label{eq:weak_anisotropy}
    \wbar \equiv \dfrac{\wx+\wy}{2} \simeq \sqrt{\wx\wy} \gg |\wy-\wx|,
\end{equation} 
where we focus on the first excited ``p-orbital'' manifold of the 2D harmonic oscillator, spanned by $\ket{\dpx}\equiv\ket{\phi_{10}}$ and $\ket{\dpy}\equiv\ket{\phi_{01}}$~\footnote{Upright p are used to distinguish the $\dpx$ and $\dpy$ orbitals from the $p_x$ and $p_y$ momenta along $x$ and $y$.} (represented in the $(x,y)$-plane in Fig.~\ref{fig:2}). The dynamics of atoms in higher orbital bands and in particular within the p-orbital manifold, have been observed \cite{isacsson_2005, Liu2006} and techniques have been developed for their controlled preparation \cite{Muller2007, Wirth_2010,Wang_2021}. Condition~\eqref{eq:weak_anisotropy} implies that the last term of Eq.~\eqref{eq:H0_1body_ladder} can be neglected, and that the subspace of $\ket{\dpx}$ and $\ket{\dpy}$ is well isolated from the rest of the one-body Hilbert space, as their energy difference $\Delta$ is much smaller than the gap $\hbar \wbar$ to other orbitals (see Fig.~\ref{fig:2}):
\begin{equation}
    \label{eq:anisotropy}
    \Delta \equiv E_{01}-E_{10} = \hbar (\omega_y-\omega_x).
\end{equation}
We define the vortex states 
\begin{equation}
    \label{eq:psi_pm}
    \ket{\psi_\pm} \equiv \dfrac{\ket{\dpx}\pm\di\ket{\dpy}}{\sqrt{2}},
\end{equation}
which, in the weakly anisotropic regime, are eigenstates of $\hL_z^{(\ob)}$ to a good approximation, with $\langle \psi_\pm | \hL_z^{(\ob)} | \psi_\pm \rangle = \pm R\hbar$ and $R \equiv \wbar/\sqrt{\wx\wy} \simeq 1$ (see Appendix~\ref{app:Lz_2D} for details). 
Their probability density $|\psi_\pm(x,y)|^2$ is shown in Fig.~\ref{fig:2}, together with their phase profiles, which reveal their opposite directions of rotation in the trap.

The ensemble of bosons in the p-orbital manifold is described by the field \mbox{$\hpsi(x,y) = \psi_+(x,y)\hb_+ + \psi_-(x,y)\hb_-$} expressed in the vortex basis, where $\hb_\pm^\dagger$ and $\hb_\pm$ are the bosonic creation and annihilation operators in the vortex mode $\pm$. In this second-quantized formalism, under the weak anisotropy criterion~\eqref{eq:weak_anisotropy}, the Hamiltonian~\eqref{eq:H0_1body_ladder} takes the form
\begin{align}
\label{eq:H_BH_0}
        \hH_0 &= \iint_{-\infty}^\infty \dd x\,\dd y\, \hpsi^\dagger(x,y)\hH_{1b}\hpsi(x,y)\\
        &= -\dfrac{\Delta}{2}\left(\hb_+^\dagger\hb_- + \hb_-^\dagger\hb_+\right) - R\hbar\Omega\left( \hn_+ - \hn_-\right) + 2\hbar\bar{\omega} \hN,\notag
\end{align}
with the number operators $\hn_\pm = \hb_\pm^\dagger\hb_\pm$ and total number operator $\hN = \hn_++\hn_-$.
The anisotropy of the trap controls the tunneling amplitude $\Delta/2$ between the vortices.

At low energy, contact interactions between cold bosons in such traps are approximated through the Hamiltonian (see Refs.~\cite{dalfovo_1999,li_2016})
\begin{equation}
\label{eq:H_BH_int}
    \begin{split}
        \hH_\dint &= \dfrac{g_{2\dD}}{2}\iint_{-\infty}^\infty \dd x\,\dd y\, \hpsi^{\dagger2}(x,y)\hpsi^2(x,y)\\
        &= -\dfrac{U}{3} \left( \hb_+^{\dagger2}\hb_+^2 + \hb_-^{\dagger2}\hb_-^2 \right) + \dfrac{2}{3}U\hN(\hN-1),
    \end{split}
\end{equation}
with $g_{2\dD}$ the 2D interaction parameter (integrating out the third dimension; see Appendix~\ref{app:bose_hubbard_model}), and $U$ the bare two-boson interaction energy. Although this contact interaction is typically repulsive ($U > 0$), it acts, in the vortex basis, as an attractive interaction between co-rotating vortices~\cite{goldman_2023}, lowering the energy by $2U/3$ per such pair. This becomes transparent by introducing the angular momentum operator acting in the many-body Hilbert space in the regime of weak anisotropy~\eqref{eq:weak_anisotropy}:
\begin{equation}
    \label{eq:Lz_MB}
    \hL_z = \iint_{-\infty}^\infty \dd x \,\dd y \, \hpsi^\dagger(x,y) \hL_z^{(\ob)} \hpsi(x,y) = R\hbar(\hn_+-\hn_-),
\end{equation}
an operator that is diagonal over the vortex basis, and in terms of which the total Bose-Hubbard Hamiltonian governing the system in the weakly anisotropic regime~\eqref{eq:weak_anisotropy} (i.e. combining Eqs.~\eqref{eq:H_BH_0} and \eqref{eq:H_BH_int}) reads
\begin{equation}
    \label{eq:H_BH}
    \begin{split}
    \hH = &\hH_0 + \hH_\dint \\
    = &-\dfrac{U}{6R^2\hbar^2}\hL_z^2 - \dfrac{\Delta}{2}\left(\hb_+^\dagger\hb_- + \hb_-^\dagger\hb_+\right)\\
    &-\Omega\hL_z+\dfrac{\hN U}{6}(3\hN-2) + 2\hbar\bar{\omega}\hN.
    \end{split}
\end{equation}
Equation~\eqref{eq:H_BH} reveals that, in the p-orbital manifold, repulsive interactions between bosons in the trap favor the population of co-rotating vortices, maximizing the squared angular momentum $\hL_z^2$. We point out that the system described here is formally equivalent to the Hamiltonian of a bosonic Josephson junction, with the two modes corresponding to vortices of opposite circulation, $\Delta$ acting as the tunneling energy, and the interaction term giving rise to nonlinear self-trapping dynamics \cite{smerzi_1997, Raghavan1999,albiez_2005,Salgueiro_2007}.

\subsection{Rotation sensing with vortex NOON states}
\label{sec:metrology}

\begin{figure}[t]
    \centering
    \includegraphics[scale=1]{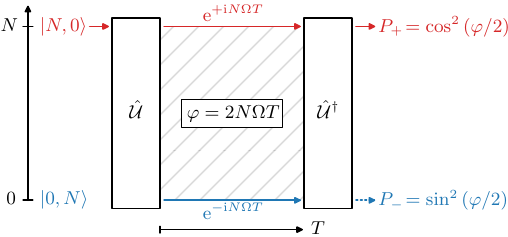}
    \caption{
    \textbf{Interferometry scheme with NOON states.}
    From left to right: The NOON operator $\hcU$ is applied to the fully-polarized initial Fock state $\ket{N,0}$, creating the NOON state. In the absence of trap anisotropy 
    ($\Delta = 0$), the NOON components $\ket{N,0}$ and $\ket{0,N}$ accumulate respectively a phase $\pm N\Omega T$ during the interrogation time $T$, amounting to the relative phase $\varphi = 2N\Omega T$ (proportional to the hatched area of the interferometer in this representation). Finally the resulting state is unprepared by the operator $\hcU^\dagger$, yielding the populations $P_+ = |\langle N,0|\psi\rangle|^2 = \cos^2(\varphi/2)$ and $P_- = |\langle 0,N|\psi\rangle|^2 = \sin^2(\varphi/2)$, from which $\varphi$ is inferred.}
    \label{fig:3}
\end{figure}

We now propose to exploit these bosonic vortices as a rotation sensor using the many-body Hamiltonian~\eqref{eq:H_BH}. The key ingredient is the last term \mbox{$-R\hbar\Omega(\hn_+ - \hn_-)$}, which imprints the effect of the external rotation $\Omega$ onto the bosonic system. As $R=\wbar/\sqrt{\wx\wy} \simeq 1$ is fixed for a given experimental setup, we absorb this coefficient into $\Omega$ in the following (see Appendix~\ref{app:Lz_2D}). Through the rotation, a Fock state $\ket{n_+, n_-}$ (describing $n_+$ and $n_-=N-n_+$ bosons occupying the vortex state $\ket{\psi_\pm}$ respectively) acquires an energy $-\hbar\Omega(n_+ - n_-)$, so that a superposition of states $\ket{\psi_a}$ and $\ket{\psi_b}$ with different population imbalances $\langle\hn_+-\hn_-\rangle_{\psi_{a,b}}$ will develop a relative phase $\varphi$ proportional to $\Omega$, that can then be measured from the interference of these states~\cite{pezze_2018}. This phase encoding is naturally optimized for $\ket{\psi_a} = \ket{N,0}$ and $\ket{\psi_b} = \ket{0,N}$, i.e. by the pure NOON state
\begin{equation}
    \label{eq:NOON}
    \ket{\dNOON} = \frac{\ket{N,0} + \de^{\di\varphi}\ket{0,N}}{\sqrt{2}}.
\end{equation}
For low anisotropy, $\Delta \ll NU$ (see Section~\ref{sec:selftrapping}), the two Fock components accumulate opposite phases $\pm N\Omega T$ over an interrogation time $T$, yielding a relative phase $\varphi = 2N\Omega T$. The full interferometric protocol is illustrated in Fig.~\ref{fig:3}: starting from the fully polarized Fock state $\ket{N,0}$ (where all the bosons rotate in the same direction) a unitary operator $\hcU$ prepares the NOON state $\ket{\dNOON} = \hcU \ket{N,0}$; after an interrogation time $T$, the inverse $\hcU^\dagger$ maps the accumulated phase back onto the population of the fully polarized Fock state $P_+ = |\langle N,0| \psi\rangle|^2$ and $P_- = |\langle 0,N| \psi\rangle|^2$, yielding
\begin{equation}
    P_\pm = \frac{1 \pm \cos(\varphi)}{2},
\end{equation}
from which $\varphi$, and thus $\Omega$, is finally inferred.

The NOON state is optimal for this task in the sense of quantum metrology~\cite{pezze_2018}. 
The rotation term $-\hbar\Omega(\hn_+-\hn_-)$ generates a unitary evolution $\de^{\di\Omega T(\hn_+-\hn_-)}$ parametrized by $\Omega T$. The quantum Fisher information $F_Q$ quantifies the sensitivity of the probe quantum state to a change in the estimated parameter (intuitively, how quickly neighboring values of $\Omega$ become distinguishable~\cite{Helstrom1976, Braunstein1994, Pezze_2014, pezze_2018}). For a pure state undergoing unitary evolution, it reduces to the variance $\sigma$ of the generator of the phase shift $\hn_+-\hn_-$: 
\begin{equation}
F_Q = 4(\sigma(\hn_+-\hn_-))^2.
\end{equation}

The NOON state, whose two Fock components differ maximally in population imbalance, maximizes this variance, yielding $F_Q= 4N^2$. The quantum Cramér-Rao bound~\cite{Rao1945, Cramer1946} then sets a lower bound on the uncertainty in the estimation of $\Omega$ when performing $\nu$ independent measurements~\cite{degen_2017,pezze_2018}:
\begin{equation}
    \label{eq:QCRB}
    \sigma\Omega \geq \frac{1}{T\sqrt{\nu F_Q}} = \frac{1}{2\sqrt{\nu}NT},
\end{equation}
which corresponds to the Heisenberg limit in its $1/N$ scaling. This is to be contrasted with the standard quantum limit $\sigma\Omega_\text{SQL} \propto 1/\sqrt{N}$, achievable with unentangled (coherent) probe states, which is optimally surpassed by the NOON state, by a factor $\sqrt{N}$~\cite{pezze_2018}.

The central experimental challenge is thus the implementation of the unitary $\hcU$ that creates the NOON state from the accessible Fock state $\ket{N,0}$~\cite{andersen_2006}. This is a non-trivial task: not only must $\hcU$ coherently transfer the population from $\ket{N,0}$ into the NOON superposition, but the NOON manifold $\{\ket{N,0}, \ket{0,N}\}$ must also remain isolated from the rest of the many-body spectrum during the interrogation time $T$ so that the phase $\varphi= 2N\Omega T$ is imprinted cleanly without leakage into other Fock states. As we show in the following subsection, both requirements are naturally met in the self-trapping regime.

\subsection{Self-trapping regime, mean-field limit and Bloch sphere}
\label{sec:selftrapping}

\begin{figure}[!t]
    \centering
    \includegraphics[scale=1]{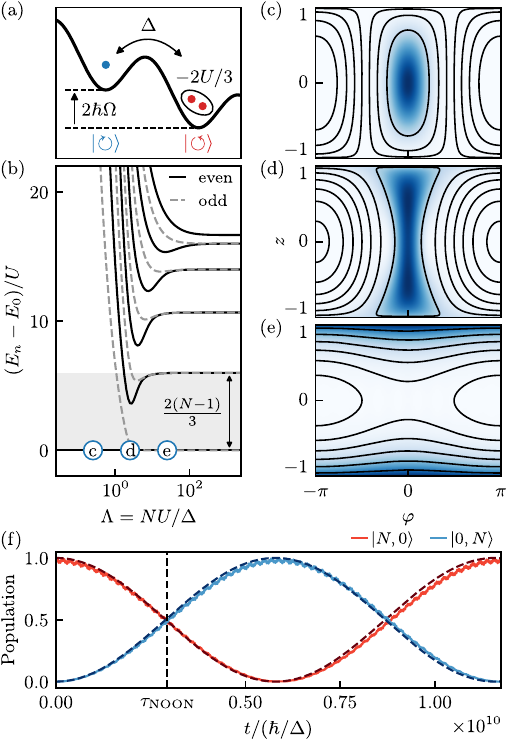}
    \caption{
    \textbf{Many-body spectrum, Bloch sphere and NOON state preparation in self-trapping regime.}
    \textbf{(a)} Sketch of the two-mode system described by the Hamiltonian~\eqref{eq:H_BH} in the vortex basis $\{\ket{\circlearrowleft}\equiv\ket{\psi_+},\ket{\circlearrowright}\equiv\ket{\psi_-}\}$~\eqref{eq:psi_pm}, with 2-body interaction energy $-2U/3$, intermode coupling $\Delta$ (i.e.~anisotropy) and energy bias $2\hbar \Omega$ (i.e.~external rotation) per bosons (colored disks).
    \textbf{(b)} Many-body excitation spectrum $(E_n-E_0)/U$ of $\hH$~\eqref{eq:H_BH} as a function of $\Lambda=NU/\Delta$ for $N=10$ and $\Omega=0$, with eigenenergies of even states in solid black and odd states in dashed gray. The self-trapping gap between the energies of $\{\ket{N,0},\ket{0,N}\}$ and $\{\ket{N-1,1},\ket{1,N-1}\}$ at large $\Lambda$ is depicted in gray.
    \textbf{(c-e)} Projection of the Bloch sphere in the $(\varphi,z)$-plane, with mean-field iso-energies (solid black lines) and Husimi representations~\cite{husimi_1940,agarwal_1981, pezze_2018} of the corresponding ground state, for $\Lambda = 0.25$ (c), 2.5 (d) and 25 (e) and $N=10$ (corresponding to the markers in panel (b)).
    \textbf{(f)} Fock populations in $\ket{N,0}$ (solid red line) and $\ket{0,N}$ (solid blue line) as a function of time, starting from $\ket{\psi_0}=\ket{N,0}$ for $\Lambda = 25$ and $N=10$, compared to the evolution of Eq.~\eqref{eq:evolution_self_trapping} (dashed lines, same color code) in the self-trapping regime with coupling coefficient Eq.~\eqref{eq:J_eff}: $\mathcal{J} = 2.71\cdot10^{-10} \Delta$. The black vertical dashed line indicates the NOON time $\tau_\dNOON=\pi\hbar/(4\mathcal{J})$, at one-quarter of this collective tunneling oscillation.
    }
    \label{fig:4}
\end{figure}

Since the two-mode Hamiltonian of Eq.~(\ref{eq:H_BH}) is formally equivalent to that of a bosonic Josephson junction with the two modes corresponding to vortices of opposite circulation, the system inherits its extensively studied dynamical regimes ~\cite{smerzi_1997, Raghavan1999,albiez_2005,anker_2005,Salgueiro_2007, Zibold2010}. In particular, we now characterize the self-trapping regime, where interactions between vortices dominate over the coupling induced by the anisotropy (see Eq.~\eqref{eq:H_BH}), and which we will propose as the operating regime for generating NOON states. In this regime, the eigenstates of Hamiltonian~\eqref{eq:H_BH} approach those of the interaction term alone (namely the Fock states $\ket{n_+, n_-}$, which have a well-defined number of bosons in each vortex mode), analogous to a two-mode realization of a Mott insulator \cite{Jaksch1998, Kuhner1998, Greiner_2002, Bloch_2008,dupont_2024}. Figure~\ref{fig:4}~(b) shows the many-body spectrum $E_n/U$ of Eq.~\eqref{eq:H_BH} as a function of the interaction parameter $\Lambda \equiv NU/\Delta$ for $N=10$ and $\Omega=0$~\footnote{In Fig.~\ref{fig:4}~(b), we plot $(E_n-E_0)/U$ versus $\Lambda=NU/\Delta$ rather than using $\Delta$ as the energy unit (as elsewhere in this work), because this choice makes the self-trapping regime immediately apparent. As $\Lambda\to\infty$, the eigenstates of $\hH$~\eqref{eq:H_BH} approach the Fock states $\ket{n_+,n_-}$, i.e. the eigenstates of the interaction term alone, whose energies $-(U/3)[n_+(n_+-1)+n_-(n_--1)-2N(N-1)]$ are linear in $U$. Measured in units of $U$, these energies become independent of $\Lambda$, so the levels flatten into horizontal lines once self-trapping sets in.}. For $\Lambda \gg 1$, the levels organize into quasi-degenerate pairs, each consisting of a symmetric and antisymmetric superposition of Fock states --- a consequence of the $+ \leftrightarrow -$ inversion symmetry of Eq.~\eqref{eq:H_BH} at $\Omega=0$. Crucially for our purposes, the two lowest eigenstates are the even and odd NOON superpositions $(\ket{N,0} \pm \ket{0,N})/\sqrt{2}$, separated from the rest of the spectrum by a gap $2(N-1)U/3$ (shaded region in Fig.~\ref{fig:4}~(b)). In the self-trapping regime, the NOON manifold is thus energetically isolated from the rest of the two-mode Hilbert space, precisely fulfilling the protection requirement identified at the end of Sec.~\ref{sec:metrology} to imprint the rotation-induced phase $\varphi$ during the interrogation time.

To gain further insight into this regime, we introduce the mean-field limit $N\rightarrow \infty$, in which quantum fluctuations around the mean values of the bosonic operators become negligible, and one makes the approximation $\hb_\pm^\dagger \rightarrow \sqrt{n_\pm}\de^{\di\theta_\pm}$, where $n_\pm$ and $\theta_\pm$ are the population and phase of the vortex mode $\ket{\psi_\pm}$. 
The relative phase $\varphi = \theta_+ - \theta_-$ and the population imbalance $z = (n_+-n_-)/N$ form a pair of canonically conjugate variables parametrizing the unit Bloch sphere of the many-body system, whose north and south poles correspond respectively to $\ket{N,0}$ and $\ket{0,N}$. The Hamiltonian~\eqref{eq:H_BH} becomes
\begin{equation}
\label{eq:H_mf}
\tilde{H} = -\dfrac{\Lambda}{3}z^2 - \sqrt{1-z^2}\cos(\varphi) - \dfrac{2\hbar\Omega}{\Delta} z,
\end{equation}
with the interaction parameter $\Lambda = NU/\Delta$, and the Hamilton equations 
\begin{equation}
    \label{eq:hamilton_eq}
    \dfrac{\dd\varphi}{\dd\tilde{t}} = \dfrac{\partial\tilde{H}}{\partial z}, \qquad \dfrac{\dd z}{\dd\tilde{t}} = -\dfrac{\partial\tilde{H}}{\partial\varphi}.
\end{equation}
In Eqs.~\eqref{eq:H_mf} and~\eqref{eq:hamilton_eq}, the tildes denote quantities expressed in units of energy $E_0 = N\Delta/2$ and time \mbox{$t_0 = \hbar/\Delta$}. This scaling reveals an effective Planck constant $\heff = \hbar/E_0t_0 =  2/N$~\footnote{Throughout this work we adopt $\heff=2/N$~\cite{lerose_2020,dupont_2026} arising from the scaling to dimensionless units above. An alternative definition $\heff=2/(N+1)$, based on state counting~\cite{lieb_1973,berezin_1975}, is used in some semiclassical analyses of bosonic systems~\cite{vanhaele_2021}; the two differ at order $1/N$ and coincide in the semiclassical limit, so the choice does not affect any conclusion of this work.}, which sets the minimal phase-space area of a many-body state on the Bloch sphere. At finite $N$, such a state $\ket{\psi}$ can be represented by its Husimi quasidistribution
of probability~\cite{husimi_1940,lieb_1973}
\begin{equation}
\label{eq:husimi}
Q(\varphi,z) = \dfrac{N+1}{4\pi}\left|\left\langle \varphi,z\,;N | \psi \right\rangle\right|^2,    
\end{equation}
proportional to the squared overlap between the state and the coherent spin state of $N$ bosons~\cite{lieb_1973,agarwal_1981}
\begin{equation}
    \label{eq:css}
    \ket{\varphi,z\,;N} = \dfrac{1}{\sqrt{N!}} \left[ \sqrt{\dfrac{1+z}{2}} \hb_+^\dagger + \de^{\di \varphi}\sqrt{\dfrac{1-z}{2}} \hb_-^\dagger \right]^N \ket{0,0},
\end{equation}
created from the vacuum $\ket{0,0}$ around $(\varphi,z)$, and with normalization $\iint Q\,\dd\varphi\dd z = 1$. Broadly speaking, $Q(\varphi,z)$ gives the probability of finding the many-body ensemble within a coherent-state patch at $(\varphi,z)$, whose surface area $4\pi/(N+1)\simeq 2\pi\heff$~\cite{Note3} vanishes as $\heff\rightarrow0$.

Figure~\ref{fig:4}~(c–e) shows iso-energy trajectories of the mean-field Hamiltonian~\eqref{eq:H_mf} (for $\Omega = 0$) for increasing values of $\Lambda$, together with the Husimi distribution of the corresponding many-body ground state at $N=10$. For $\Lambda \ll 1$ (Fig.~\ref{fig:4}~(c)), the iso-energies form closed orbits around $(\varphi,z)=(0,0)$ and $(\pi,0)$, and the ground state concentrates there. As $\Lambda$ increases (Fig.~\ref{fig:4}~(c) to (e)), the phase space separates into two disjoint sets of orbits trapped in the northern $(z>0)$ and southern $(z<0)$ hemispheres --- the classical signature of self-trapping~\cite{smerzi_1997, albiez_2005, anker_2005} --- and the ground-state Husimi distribution splits into two lobes located at the poles, i.e. near $\ket{N,0}$ and $\ket{0,N}$. Remarkably, the Husimi distribution closely follows the mean-field iso-energies already for $N=10$, confirming the relevance of the mean-field description even at moderate particle number. For the remainder of this work, we focus on the self-trapping regime, which, combined with the relatively weak-interaction condition ensuring confinement to the $p$-band manifold (Sec.~\ref{sec:bh_model}), defines the energy hierarchy of our study
\begin{equation}
\label{eq:hierarchy}
\Delta \ll NU \ll \hbar\bar\omega.
\end{equation}

In the self-trapping regime, the isolation of the NOON manifold $\{\ket{N,0}, \ket{0,N}\}$ from the rest of the spectrum allows the dynamics within this subspace to be captured by an effective two-level Hamiltonian. Since the interaction term of Hamiltonian~\eqref{eq:H_BH} is diagonal in the Fock basis and identical for $\ket{N,0}$ and $\ket{0,N}$, it contributes to a global phase that can be omitted, leaving only the rotation and anisotropy-induced coupling:
\begin{equation}
\label{eq:hred}
\hH_{\text{red}} = \begin{pmatrix}
-N\hbar\Omega & -\mathcal{J}  \\ 
-\mathcal{J} & N\hbar\Omega \\
\end{pmatrix},
\end{equation}
with an effective coupling (see Appendix~\ref{app:perturbation_theory})
\begin{equation}
\label{eq:J_eff}
    \mathcal{J} \approx \frac{N\Delta}{2(N-1)!}\left(\dfrac{3\Delta}{4U}\right)^{N-1}.
\end{equation}
This reduced description highlights a striking feature of the self-trapping regime: starting from the fully polarized Fock state $\ket{N,0}$, the system undergoes a collective tunneling oscillation to $\ket{0,N}$. For $\Omega =0$, this evolution reads
\begin{equation}
\label{eq:evolution_self_trapping}
    \ket{\psi(t)} = \cos\left(\dfrac{\mathcal{J}t}{\hbar}\right) \ket{N,0} + \di \sin\left(\dfrac{\mathcal{J}t}{\hbar}\right) \ket{0,N},
\end{equation}
and at time $\tau_\dNOON = \pi\hbar/(4\mathcal{J})$, the system reaches the equal superposition $(\ket{N,0} +\di \ket{0,N})/\sqrt{2}$,  directly realizing the unitary $\hcU$ required by the interferometric protocol of Sec.~\ref{sec:metrology}. Figure~\ref{fig:4}~(e) shows the excellent agreement between the exact many-body dynamics and the reduced evolution~\eqref{eq:evolution_self_trapping} for $N=10$ and $\Lambda = 25$. However, the exponential suppression of $\mathcal{J}$ with $N$ translates into a prohibitively long NOON preparation time. For $N=10$ and $U/\Delta=2.5$, one finds $\tau_\dNOON = 2.93\cdot10^9$~$\hbar/\Delta$, corresponding to ~$1.7\cdot10^7$~s for realistic experimental parameters (see Appendix~\ref{app:exp_param}), far exceeding the coherence times accessible on current cold-atom platforms. 

This exponentially low coupling between the NOON Fock states motivates the search for alternative preparation schemes, two of which we present in the remainder of this paper.

\section{Accelerated creation of vortex NOON states}
\label{sec:results_of_accelerated_preparations}

To steer the system towards a NOON state, we consider a control over the energy difference between the two vortex modes, implemented through the rotation $\Omega(t)$ of the trap. The Bose-Hubbard Hamiltonian describing the system thus takes the form:
\begin{align}
\label{eq:BHMomega}
    \notag\hH = &-\dfrac{U}{3} \left( \hb_+^{\dagger2}\hb_+^2 + \hb_-^{\dagger2}\hb_-^2 \right) -\frac{\Delta}{2} \left(\hb_+^\dagger\hb_- +\hb_-^\dagger \hb_+\right)  \\
    & - \hbar\Omega(t) (\hn_+ - \hn_-),
\end{align}
where $\Omega(t) = \Omega_\text{rot} + \Omega_\text{control}(t)$ comprises the small rotation $\Omega_\text{rot}$ to be measured and the applied control $\Omega_\text{control}(t)$, which acts only during the NOON preparation. The two are vastly separated in scale: the target rotations are tiny (e.g. the Earth's rotation $\hbar\Omega_\oplus \sim 10^{-7}\,\Delta$; see Appendix~\ref{app:exp_param}), whereas the control amplitudes used here are of order $\hbar\Omega_\text{control}\sim1$ to $10\,\Delta$. What matters, however, is not this ratio alone but the phase $2N\Omega_\text{rot}\tau_\dNOON$ that the rotation imprints over the preparation time (see Sec.~\ref{sec:metrology}). The accelerated protocols presented in this section make $\tau_\dNOON$ short enough that this phase remains far below the $\pi/2$ phase rotation of the collective NOON state during preparation~\footnote{For the Earth's rotation and the preparation times reached here ($\tau_\dNOON \sim 10^2$ to $10^3\,\hbar/\Delta$), $2N\Omega_\oplus\tau_\dNOON\sim(10^{-5}-10^{-3})\ll\pi/2$ (for $N$ ranging from 3 to 10).}. We therefore neglect $\Omega_\text{rot}$ during the preparation (an approximation motivated in Appendix~\ref{app:exp_param}.)

To assess and compare the two protocols developed in this section, we quantify their performance through the NOON-state fidelity of the produced state $\ket{\psi}$:
\begin{equation}
    \label{eq:fidelity}
    \mathcal{F}_\dNOON = \left| \langle \dNOON | \psi \rangle \right|^2,
\end{equation}
which equals unity for the ideal target state of Eq.~\eqref{eq:NOON}~\footnote{With $\varphi$ in Eq.~\eqref{eq:NOON} the phase produced by the protocol.}, otherwise capturing the imperfections of a realistic, finite-time preparation. Together with the preparation time, $\mathcal{F}_\dNOON$ serves as the central figure of merit throughout this section. For small particle numbers ($N\leq8$), we develop an adiabatic driving protocol that steers the system towards the vortex NOON state (Sec.~\ref{sec:gcd}). For larger particle numbers ($N > 8$), the system enters a semiclassical regime where resonance- and chaos-assisted tunneling yields high fidelities on experimentally feasible timescales (Sec.~\ref{sec:RATCAT}). We note that our preparation schemes assume loading into the p-state manifold, which can be achieved using the techniques of Refs.~\cite{isacsson_2005, Liu2006,Muller2007, Wirth_2010,Wang_2021}.

\subsection{Creation of vortex NOON states from counterdiabatic driving}
\label{sec:gcd}

In this section, we realize the NOON-preparation unitary $\hcU$ (see Fig. \ref{fig:3}) through adiabatic driving of the control $\Omega(t)$ acting on the energy difference between the vortex modes (Eq.~\eqref{eq:BHMomega}), which drives the system along the instantaneous eigenstates of the Hamiltonian~\eqref{eq:BHMomega}.

\subsubsection{Shortcut to adiabaticity}

The possibility of adiabatically generating a NOON state is opened by the separation of the energy levels corresponding to the states of interest. The adiabatic theorem states that when the Hamiltonian varies slowly compared to the intrinsic timescale set by the level splitting between the energy eigenstates, the system remains in the instantaneous eigenstate in which it was initially prepared. The driving can therefore be optimized to vary slowly in the regions where diabatic transitions are likely, i.e. where the energy levels are closest, and faster elsewhere, where the levels are sufficiently separated such that transitions towards undesired states can be neglected. Several standard choices of driving functions exist and are commonly used~\cite{Landau_1932,Zener_1932,Majorana_1932,Stuck_1932}. However, it is possible to determine the optimal path in parameter space, minimizing the protocol duration~\cite{tomka_2016,Carlini_2006, Nauth_2022}. To do so, one must understand the geometry of the system encoded in the Fubini-Study metric, which describes the distance between two quantum states in Hilbert space as a function of the time-dependent parameter~\cite{Provost_1980}. To find the optimal $\Omega(t)$ to apply to the system to steer it adiabatically, one has to solve the following equation defining geodesics in parameter space \cite{Misner1973, tomka_2016}:
\begin{equation}
\label{eq:geodesic_equation}
    \hbar^2 g_{\Omega\Omega}\dot{\Omega}^2=\mathrm{const.}
\end{equation}
where
\begin{equation}
\label{eq:metric_tensor}
    g_{\Omega\Omega} = \text{Re}\sum_{m\neq n}\frac{\langle n\vert \partial_\Omega \hH_{\text{red}}\vert m\rangle\langle m\vert \partial_\Omega \hH_{\text{red}}\vert n\rangle }{(E_n - E_m)^2},
\end{equation}
is the only non-vanishing component of the metric tensor~\cite{Misner1973}. Here, $\ket{n}$ and $\ket{m}$ are the two instantaneous eigenstates of the reduced Hamiltonian~\eqref{eq:hred} at $\Omega$, with eigenenergies $E_n$ and $E_m$~\footnote{For the two-level reduced system of Eq.~\eqref{eq:hred}, the sum~\eqref{eq:metric_tensor} reduces to a single term.}. Equation~\eqref{eq:geodesic_equation} gives a condition on how $\Omega(t)$ should behave for the system to follow the optimal path, minimizing at each time the local infidelity, i.e. the distance between its current state and the target eigenstate.

For the reduced Hamiltonian~\eqref{eq:hred}, the spectrum can be computed exactly and the geodesic condition yields the optimal time parametrization for $\Omega$ \cite{tomka_2016, Dengis_2025_b}:
\begin{equation}
\label{eq:geodesic}
    N\hbar\Omega (t) = \mathcal{J}\tan(\alpha_i + (\alpha_f-\alpha_i)\,t/T),
\end{equation}
where $\alpha_{i,f} = \tan^{-1} (N\hbar\Omega_{i,f}/\mathcal{J})$ are taken close to $\pm \pi/2$ so as to approach the tangent asymptote and ensure that the states $\ket{N,0}$ and $\ket{0,N}$ are initially well decoupled. This requires choosing $\Omega_i \equiv \Omega(t=0)$ and $\Omega_f \equiv \Omega(t=T)$ such that $\left|N\hbar\Omega_{i,f}/\mathcal{J}\right| \gg 1$, thereby placing the tangent function close to its asymptotic regime. This function prescribes a fast evolution in the regions far from the central gap, and a very slow evolution near the high-risk zones for diabatic transitions, where the energy levels are closest.

While the geodesic parametrization~\eqref{eq:geodesic} optimally guides the system along the instantaneous eigenstates, the adiabatic theorem still constrains its evolution to remain slow. This speed limitation can be overcome by introducing a counterdiabatic Hamiltonian (CDH), which compensates for the system’s inertia and allows one to manipulate the state as quickly as desired~\cite{Berry_2009}. In the eigenbasis of the Bose-Hubbard Hamiltonian, the CDH takes the form
\begin{equation}
\label{eq:hcd}
    \hH_{\text{CD}}(t) = i\hbar \sum_{n}\sum_{m\neq n} \frac{\langle m \vert\dot{\hH}(t)\vert n \rangle}{E_{n}-E_{m}}\vert m \rangle \langle n \vert.
\end{equation}
Thus, defining a new Hamiltonian $\hH + \hH_{\text{CD}}$ cancels the diabatic couplings in the eigenbasis, protecting the system state from unwanted transitions between instantaneous eigenstates. While it is often difficult to construct the CDH for large-scale systems, approximations exist to circumvent the need for full knowledge of the exact spectrum~\cite{Claeys_2019}, or to select only the matrix elements relevant to the task~\cite{Petiziol_2018}. In the approach we employ, the two-level form of the reduced Hamiltonian~\eqref{eq:hred} enables an exact calculation of the CD term required for the creation of NOON states~\cite{Dengis_2025_b}:
\begin{equation}
    \label{eq:hcd_explicit}
    \hH_{\text{CD}}(t) = \frac{\mathcal{J}N\hbar^2\dot{\Omega}(t)}{2\left(\mathcal{J}^2 + (N\hbar\Omega(t))^2\right)}\,\hat{\sigma}_y,
\end{equation}
with $\hat{\sigma}_y$ the Pauli matrix. The matrix elements of the counterdiabatic Hamiltonian are thus purely off-diagonal and imaginary. Since its purpose is to compensate for the inertia induced by excessively fast driving, it is proportional to the driving rate $\dot{\Omega}$ and inversely proportional to the squared gap $\mathcal{J}^2 + (N\hbar\Omega(t))^2$.

\subsubsection{Geodesic counterdiabatic driving}

A priori, the CDH is defined independently of the choice of the driving function $\Omega$. However, important connections have been established between geodesic and counterdiabatic driving \cite{delCampo_2012,Kolodrubetz_2013,Dengis_2025_a,Dengis_2025_b}. The combination of these two driving strategies, known as geodesic counterdiabatic driving (GCD), makes it possible to render the counterdiabatic matrix element time-independent, which is clearly advantageous for an experimental implementation of the protocol. Indeed, the purpose of geodesic driving is precisely to suppress variations near the gap, where the matrix elements of the CDH become significant. Conversely, the elements of the CDH are nearly zero where $\dot{\Omega}$ becomes very large. By inserting the expression of the geodesic driving~\eqref{eq:geodesic} into the definition of the CDH~\eqref{eq:hcd_explicit}, one obtains the constant amplitude of the GCD protocol:
\begin{equation}
\label{eq:omega}
    \frac{\mathcal{J}N\hbar^2\dot{\Omega}(t)}{2\left(\mathcal{J}^2 + (N\hbar\Omega(t))^2\right)} = \frac{(\alpha_f -\alpha_i)\hbar}{2T},
\end{equation}
The norm of the counterdiabatic term is therefore fully determined by the total evolution time $T$, since \mbox{$|\alpha_f - \alpha_i| \approx \pi$} (see Eq.~\eqref{eq:geodesic}). Under the GCD protocol, the effective gap becomes $\sqrt{\mathcal{J}^2 + (N\hbar\Omega(t))^2 + (\pi\hbar/2T)^2}$, broadened by the last term, which is inversely proportional to $T$. Hence, the GCD NOON time at resonance ($\Omega=0$) for a total evolution time $T \ll \hbar/\mathcal{J}$ is given by
\begin{equation}
    \tau_\dNOON = \frac{\pi\hbar}{4\sqrt{\mathcal{J}^2 + (\pi\hbar/2T) ^2}} \approx T/2.
\end{equation}
The result of Eq.~\eqref{eq:omega} is intrinsically connected to the notion of a quantum speed limit \cite{Mandelstam_1945, Deffner_2014, Dengis_2025_b}. Indeed, the relation linking counterdiabatic control to geodesic driving arises through the diagonal elements of the CDH:
$\langle n | \hH_{\text{CD}}^{2} | n \rangle = \hbar^{2} g_{\Omega\Omega} \dot{\Omega}(t)^{2}$.
In our case, and more generally for a two-level system, the squared norm of the CDH can therefore be directly related to the geodesic condition, which requires that the quantity $\hbar^2g_{\Omega\Omega}\dot{\Omega}(t)^{2} = \Delta E^2$ remain constant in time~\cite{delCampo_2012, Kolodrubetz_2013, Kolodrubetz_2017}. Thus, by linking $(\pi\hbar/2T)^2$ to $\Delta E^2$ and invoking the Mandelstam-Tamm bound for the system under consideration~\cite{Mandelstam_1945}, one finds that the GCD protocol saturates the quantum speed limit by virtue of the expression~\eqref{eq:omega}.

Thus, when the GCD protocol is applied to the system, the gap is broadened by the amplitude of the CDH, which is inversely proportional to the total time $T$. To apply the protocol to the full Bose-Hubbard Hamiltonian~\eqref{eq:BHMomega}, one must define an effective complex hopping that incorporates the action of the counterdiabatic term. Using the identification 
\begin{equation}
\label{eq:identification}
    \hH_{\text{red}}(U,\Delta,\Omega(t)) + \hH_{\text{CD}} \stackrel{!}{=} \hH_{\text{red}}\left(U,\Delta_{\text{eff}},\Omega(t)\right),
\end{equation}
we define a complex $\Delta_{\text{eff}}$ such that
\begin{equation}
\label{eq:jeff}
    \Delta_{\text{eff}}/2 = \left[ (\Delta /2)^N - i\frac{\pi\hbar (2U/3)^{N-1}(N-1)!}{2NT} \right]^{1/N}.
\end{equation}
The practical engineering of this term will be described in Sec.\ref{sec:FMBHM}. In the self-trapping regime, the Hamiltonian describing the system under the GCD protocol is given by 
\begin{align}
\label{eq:BHMGCD}
    \notag\hH = &-\dfrac{U}{3} \left( \hb_+^{\dagger2}\hb_+^2 + \hb_-^{\dagger2}\hb_-^2 \right) -\frac{1}{2} \left(\Delta_{\text{eff}} \,\hb_+^\dagger\hb_- + \Delta_{\text{eff}}^*\,\hb_-^\dagger \hb_+\right)  \\
    &- \hbar\Omega(t) (\hn_+ - \hn_-),
\end{align}
which describes an adiabatic time evolution along the eigenvectors of Hamiltonian~\eqref{eq:BHMomega}, requiring only a time-dependent driving function.

\begin{figure}[!t]
    \centering
    \includegraphics[width=0.5\textwidth]{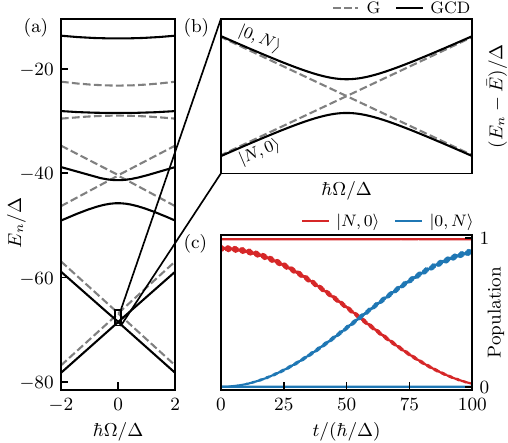}
    \caption{
    \textbf{Geodesic counterdiabatic driving for fast NOON-state preparation.}
    \textbf{(a)} Spectrum of the Bose-Hubbard Hamiltonian~\eqref{eq:BHMomega} under geodesic driving (G, dashed gray) and geodesic counterdiabatic driving (GCD, solid black) as a function of the bias $\hbar\Omega/\Delta$, for $N=5$ particles, $U=5\,\Delta$, and preparation time $T=10^2\,\hbar/\Delta$. The two lowest levels host the NOON manifold $\{\ket{N,0},\ket{0,N}\}$, isolated from the rest of the spectrum by the reduced self-trapping gap $\sim NU/\Delta$.
    \textbf{(b)} Zoom on the avoided crossing between $\ket{N,0}$ and $\ket{0,N}$ at $\Omega=0$ for the two protocols (translated by $\bar{E} = (E_0+E_1)/2$). The GCD avoided crossing is significantly broadened with respect to G, enabling a faster adiabatic evolution.
    \textbf{(c)} Time evolution of the Fock populations in $\ket{N,0}$ (red) and $\ket{0,N}$ (blue) for a total time $T=10^2\,\hbar/\Delta$. GCD achieves a full population inversion, creating the NOON state near the midpoint $t\approx T/2$, whereas geodesic driving alone (G) fails to invert the population.
    }
    \label{fig:populationsgcd}
\end{figure}

To illustrate the effectiveness of the GCD technique for the on-demand creation of a NOON state~\eqref{eq:NOON}, we apply the protocol to the population inversion of the system described by Hamiltonian (\ref{eq:BHMGCD}). We stress that, while the GCD ramps result from analytical arguments for $\hH_\text{red}$~\eqref{eq:hred}, all of the simulations we present were performed within the framework of the complete $N$-body system, going beyond the effective two-level reduction. We consider the fully polarized initial Fock state $\ket{\psi_0}=\ket{N,0}$, with the two modes separated by a bias $2N\hbar\Omega$. This bias is then varied adiabatically in time according to the GCD protocol. At the midpoint of the evolution, an avoided crossing is reached and the population is inverted adiabatically, creating the NOON state.

Figure~\ref{fig:populationsgcd}~(a) shows the effect of the GCD protocol by comparing the spectra of Hamiltonians~\eqref{eq:BHMomega} (dashed gray) and~\eqref{eq:BHMGCD} (solid black). The GCD protocol increases the separation between the energy levels of the NOON subspace at the avoided crossings. The relevant states, and specifically the avoided crossing between $\ket{N,0}$ and $\ket{0,N}$, are detailed in the zoom of Fig.~\ref{fig:populationsgcd}~(b). The inset highlights the avoided crossing that appears under driving without counterdiabatic terms, and compares it with that of the GCD protocol, which is visibly broadened, enabling a faster adiabatic evolution. Figure~\ref{fig:populationsgcd}~(c) shows the population dynamics over time for a total protocol duration of $T = 10^2\,\hbar/\Delta$. The GCD method enables a full population inversion, whereas using geodesic driving alone achieves roughly $1\%$ inversion. The NOON state is generated halfway through the protocol, i.e., at $T\approx50\,\hbar/\Delta$.

\begin{figure}[!t]
    \centering
    \includegraphics[scale=1]{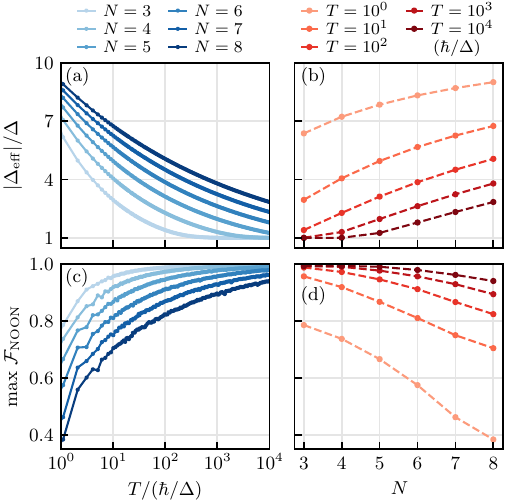}
    \caption{
    \textbf{Performance of the GCD with protocol duration and particle number.}
    \textbf{(a,b)} Magnitude of the effective coupling $|\Delta_\mathrm{eff}|$~\eqref{eq:jeff} as a function of (a) the total protocol duration $T$ (in units of $\hbar/\Delta$) for different particle numbers $N$, and (b) the particle number $N$ for different total durations $T$. 
    \textbf{(c),(d)} Maximum fidelity $\mathcal{F}_\mathrm{NOON}$~\eqref{eq:fidelity} reached during the evolution toward the NOON state as a function of (c) the total protocol duration $T$ for different particle numbers, and (d) the particle number $N$ for different total durations. The GCD protocol generates a near-unit-fidelity NOON state for $N=5$ in about $T\approx 10^2\,\hbar/\Delta$ for $NU=25\Delta$.}
    \label{fig:fidelity}
\end{figure}

The broadening of the levels induced by the GCD comes at a cost visible in Fig.~\ref{fig:populationsgcd}~(a): the larger effective coupling $|\Delta_\deff|$ that widens the avoided crossing between $\vert N,0\rangle$ and $\vert 0,N\rangle$ simultaneously narrows the self-trapping gap separating the NOON manifold from the neighboring Fock states. The GCD spectrum (solid black) thus exhibits a smaller self-trapping gap than the geodesic one (dashed gray). As long as this isolating gap remains open, the two-level reduction of Eq.~\eqref{eq:hred} holds and the dynamics stays confined to the NOON manifold. Its progressive closing at larger $|\Delta_\deff|$ constitutes the ultimate limitation of this approach. This is clearly seen in Fig.~\ref{fig:fidelity}, whose top panels show the magnitude of the effective hopping amplitude $\Delta_\mathrm{eff}$ as a function of (a) the total time $T$ and (b) the number of particles $N$. The faster the state-preparation protocol is performed, the larger are the experimental resources that are required to implement the effective anisotropy. Since the latter scales inversely with $T$ (see Eq.~\eqref{eq:jeff}), choosing a total time that is too short (see Eq.~(\ref{eq:hierarchy})) may drive the system out of the self-trapping regime through an insufficient ratio between the interaction strength and the norm of the anisotropy.

The bottom panels of Figure~\ref{fig:fidelity} show the maximum fidelity obtained during the evolution as a function of (c) the total time $T$ and (d) the number of particles $N$. The fidelity is limited by the tradeoff identified in Fig.~\ref{fig:populationsgcd}~(a): accelerating the protocol requires a large effective coupling $|\Delta_\deff|$ (Fig.~\ref{fig:fidelity}~(a,b)), which narrows the self-trapping gap and lets population leak out of the NOON manifold $\{\ket{N,0},\ket{0,N}\}$ into neighboring Fock states. Since the bare coupling $\mathcal{J} \sim (\Delta/2)^N/(N-1)!$ collapses faster than exponentially with $N$ (Eq.~\eqref{eq:J_eff}), maintaining a fixed preparation time at larger $N$ requires an even larger $|\Delta_\deff|$, so this leakage sets in earlier. As a result, there is a minimum total evolution time required before a sufficiently high fidelity can be achieved, and this time grows with the number of particles : near unit fidelity is reached in $T\sim10^2$ to $10^3\,\hbar/\Delta$ for $N=5$ particles, but requires $T\gtrsim 10^4\,\hbar/\Delta$ for $N=8$. The GCD protocol enables very high fidelities to be reached at short times for a small number of particles, but becomes progressively less effective as $N$ increases. The practical ceiling of this approach lies near $N\approx8$ for the parameters of this study, and motivates the semiclassical route of Sec.~\ref{sec:RATCAT}. 

\subsubsection{Floquet-Magnus emulation of the complex coupling}
\label{sec:FMBHM}
The need to introduce a complex coupling $\Delta_\deff$ may, however, pose practical difficulties. A viable solution to implement $\Delta_\deff$ is through Floquet engineering, by adding an oscillating term to the Bose-Hubbard Hamiltonian, emulating the additional complex phase~\cite{Goldman_2014, Petiziol_2018, Claeys_2019,Petiziol_2024a}. We show that such an alternative construction is possible at the cost of micro-motion at the Floquet frequency in the population dynamics over time. 

Let us consider a Bose-Hubbard Hamiltonian as defined in Eq.~\eqref{eq:BHMomega}, to which we add an oscillating term
\begin{equation}
\label{eq:FM}
    \hat{h}(t) = A\sin(\omega t)(\hat{b}_+^\dagger \hat{b}_- + \hat{b}_-^\dagger \hat{b}_+) + B\cos(\omega t)(\hat{b}_+^\dagger \hat{b}_+-\hat{b}_-^\dagger \hat{b}_-),
\end{equation}
parametrized by the amplitudes $A$ and $B$. At first order of the Floquet-Magnus expansion, the oscillating terms (\ref{eq:FM}) lead to an effective static contribution to the Hamiltonian~\cite{magnus_1954,rahav_2003,Goldman_2014}, 
\begin{equation}
\label{eq:FM:firstorder}
    \hat{h}_\deff = \di (\hbar\omega)^{-1}AB\left( \hat{b}_+^\dagger \hat{b}_- - \hat{b}_-^\dagger \hat{b}_+\right) + \mathcal{O}((\hbar\omega)^{-2}),
\end{equation}
which allows us to identify, by comparison with the Hamiltonian~\eqref{eq:BHMGCD}, the complex term to be emulated
\begin{equation}
\label{eq:FM_identification}
    AB/(\hbar\omega) = \text{Im}\left\{\Delta{_{\text{eff}}} \right\}.
\end{equation}
Fixing $B=A$, we obtain the condition
\begin{equation}
\label{eq:FM_amplitude}
    A = \sqrt{\hbar\omega\,\vert\,\text{Im}\left\{\Delta{_{\text{eff}}} \right\}\vert}.
\end{equation}
The frequency $\omega \equiv \omega_{\text{F}}$ should ideally be sufficiently large compared to the characteristic frequencies of the system. We therefore define three characteristic frequency scales:
the adiabatic frequency $\omega_{\text{GCD}} \sim \sqrt{\mathcal{J}^2 + (\pi\hbar/2T)^2}/\hbar \sim 1/T$, which sets the rate at which the population is adiabatically inverted; 
the many-body frequency $\omega_{\text{MB}} \sim 2U(N-1)/(3\hbar)$, arising from the energy difference between the states $\ket{N,0}$ and $\ket{N-1,1}$;
and finally the Floquet frequency $\omega_{\text{F}}$. 
It is necessary that $\omega_{\text{GCD}}$ be the smallest frequency in the system, despite the acceleration of the adiabatic process enabled by the GCD protocol. The excitation frequency to the nearest many-body states, $\omega_{\text{MB}}$, must be sufficiently large to remain decoupled from the adiabatic driving (in other words, to stay in the self-trapping regime). The Floquet frequency $\omega_{\text{F}}$ must be well separated from the other system frequencies to avoid unwanted couplings. Finally, all internal frequencies must remain sufficiently small compared to the average trapping frequency, which isolates the $p$ band from the $s$ and $d$ bands of the one-body spectrum (see Sec.~\ref{sec:bh_model}). These requirements can be expressed as
\begin{equation}
\label{eq:C5_hierarchy}
    \omega_{\text{GCD}}\ll \omega_{\text{MB}}\ll \omega_{\text{F}} \ll \bar\omega,
\end{equation}
ensuring that the different timescales at which the relevant physical processes occur remain well separated.

\begin{figure}[t]
    \centering
    \includegraphics[scale=1]{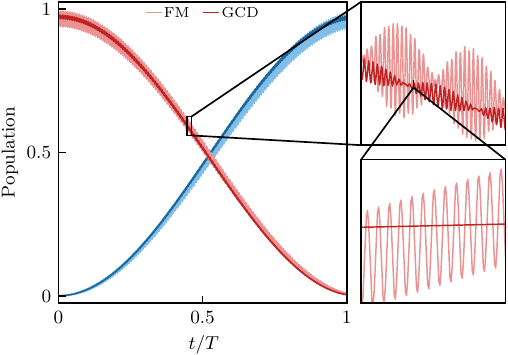}
    \caption{
    \textbf{Floquet-Magnus emulation of the complex hopping.}
    Time evolution of the populations in $\ket{N,0}$ (red) and $\ket{0,N}$ (blue) for geodesic counterdiabatic driving (Eq. ~\eqref{eq:BHMGCD}) (dark colors) and for its emulation by a Hamiltonian with oscillating terms (Eq.~\eqref{eq:FM}) (light colors), with $N=5$, $T=10^{3}\,\hbar/\Delta$, $NU=25\,\Delta$ and $\omega_\mathrm{F}=2\cdot10^{4}\,\Delta/\hbar$. The two methods share the same average dynamics, reproducing an adiabatic population inversion without resorting to a complex hopping.
    The successive zooms resolve the three characteristic frequencies of the system: $\omega_\mathrm{GCD}$, set by the adiabatic evolution (first panel); $\omega_\mathrm{MB}$, associated with many-body transitions to neighboring Fock states (second panel); and $\omega_\mathrm{F}$, related to the Floquet-Magnus emulation of the complex hopping (third panel). Their separation in magnitude (Eq.~\eqref{eq:C5_hierarchy}) ensures that they do not interfere.
    }
    \label{PopulationsJXJZ}
\end{figure}

As we showed, it is possible to emulate the GCD protocol through additional oscillating terms. Figure~\ref{PopulationsJXJZ} compares the population dynamics obtained using the Bose-Hubbard Hamiltonian with a complex coupling $\Delta_{\text{eff}}$ (dark curves) and using the oscillating terms added via the Hamiltonian~\eqref{eq:FM} (light curves). In a parametric regime where the Floquet frequency is sufficiently large to be fully decoupled from the other characteristic frequencies of the system, the main dynamical features produced by the two methods are the same. However, larger oscillations arise and the evolution is governed by $\hat{H}(U,\Delta) + \hat{h}(t)$. These oscillations resolve the different frequencies at different scales, as shown in the insets. The first zoom of Fig.~\ref{PopulationsJXJZ} reveals oscillations arising from many-body couplings with other states in the spectrum. A beating phenomenon is observed, corresponding to the proximity between the excitation frequencies connecting neighboring states. Denoting by $E_{n_+,n_-}$ the energy of the state with $n_\pm$ bosons in the $(\pm)$ rotating mode \footnote{These are close to the energies of the Fock states, but are slightly modified by the anisotropy $\Delta$ (see Appendix \ref{app:perturbation_theory})}, the two frequencies are
\begin{equation}
\begin{split}
    &\hbar\omega_{\text{MB1}} = E_{N,0}-E_{N-1,1}\\
    &\hbar\omega_{\text{MB2}} = E_{N-1,1}-E_{N-2,2}
\end{split}
\end{equation}
The proximity of these two frequencies gives rise to the observed beating, visible in both curves, through the ratio
$\omega_{\text{MB1}}/\omega_{\text{MB2}}$. The second zoom of Fig.~\ref{PopulationsJXJZ} shows the third and largest frequency, corresponding to the Floquet-Magnus emulation of the complex coupling. These results show that once the three main frequencies of the system are sufficiently well separated in magnitude, they no longer interfere with the physical processes associated with each of them.

Overall, it is worth emphasizing that the protocol developed here is particularly effective for the fast generation of vortex NOON states with a small number of particles. Indeed, for increasing numbers of bosons, the time required to create a NOON state with good fidelity through GCD increases, as does the required magnitude of $|\Delta_\mathrm{eff}|$ (Fig.~\ref{fig:fidelity}), which eventually drives the operating point out of the self-trapping regime (Sec.~\ref{sec:selftrapping}). For larger particle numbers, we next propose an alternative route to generate vortex entanglement based on resonance- and chaos-assisted tunneling.

\subsection{Creation of vortex NOON states from resonance- and chaos-assisted tunneling}
\label{sec:RATCAT}

For larger numbers of bosons $N$, where the geodesic counterdiabatic-driving approach of Sec.~\ref{sec:gcd} becomes impractical, the system enters a semiclassical regime in which the mean-field description introduced in Sec.~\ref{sec:selftrapping} becomes increasingly accurate. In this regime, a periodic modulation of a parameter of the Hamiltonian generically renders the classical dynamics non-integrable, producing a mixed regular-chaotic phase space~\cite{bohigas_1993} whose structure can be exploited to dramatically accelerate the NOON-state preparation through the semiclassical phenomena of resonance- and chaos-assisted tunneling (RAT and CAT)~\cite{tomsovic_1994,Brodier_2001, eltschka_2005,schlagheck_2011,vanhaele_2021,vanhaele_2022}. The next two Sections~\ref{subsub:driven_ratcat} and \ref{subsub:ratcat} provide a detailed introduction to the RAT and CAT effects in a many-body context. The final Section~\ref{subsub:ratcat_results} presents a concrete example of the accelerated preparation of a vortex NOON state using RAT and CAT.

\subsubsection{Driven two-mode system}
\label{subsub:driven_ratcat}

Assuming, as before, control over the energy difference between the vortex modes $\pm$, we consider the following periodic drive of the Hamiltonian~\eqref{eq:BHMomega}:
\begin{align}
    \label{eq:H_drive}
    \hH(t) = &-\dfrac{U}{3} \left( \hb_+^{\dagger2}\hb_+^2 + \hb_-^{\dagger2}\hb_-^2 \right) - \dfrac{\Delta}{2} \left( \hb_+^\dagger\hb_- + \hb_-^\dagger \hb_+ \right) \notag\\
    & - \hbar\Omega \cos(\omega t) (\hn_+ - \hn_-),
\end{align}
with $\hbar\Omega$ and $\omega$ the amplitude and angular frequency of the modulation. By analogy with Bloch's theorem in space, time-periodic Hamiltonians are 
conveniently treated within the Floquet formalism~\cite{eckardt_2017}. We denote by $\hU_\dF$ the Floquet evolution operator over one driving period and by $\{\ket{\phi_j}\}$ its eigenstates:
\begin{equation}
    \label{eq:floquet}
    \hU_\dF \ket{\phi_j} = \de^{-\di 2\pi\varepsilon_j/\hbar \omega}\ket{\phi_j},
\end{equation}
with quasienergies $\varepsilon_j$ defined modulo $\hbar \omega$~\cite{grifoni_1998,eckardt_2017}. Although $\hH(t)$~\eqref{eq:H_drive} is not invariant under the mode exchange \mbox{$\hR: \hb_+ \leftrightarrow \hb_-$}, it commutes with the combined operation $\hPi \equiv \hR\,\hT_{T/2}$ that exchanges modes and shifts time by half a period (with $T=2\pi/\omega$). This generalized parity~\cite{grifoni_1998} classifies the Floquet states into even ($+$) and odd ($-$) sectors and protects the existence of a NOON-supporting doublet $\ket{\phi_\pm}$ in the self-trapping regime, $\hPi \ket{\phi_\pm} = \pm \ket{\phi_\pm}$, whose quasienergy splitting $\Delta\varepsilon_\pm=|\varepsilon_+-\varepsilon_-|$ governs the NOON time as 
\begin{equation}
    \label{eq:NOON_time_drive}
    \tau_\dNOON = \dfrac{\pi\hbar}{2\Delta\varepsilon_\pm}.
\end{equation}

\begin{figure}[!t]
    \centering
    \includegraphics[scale=1]{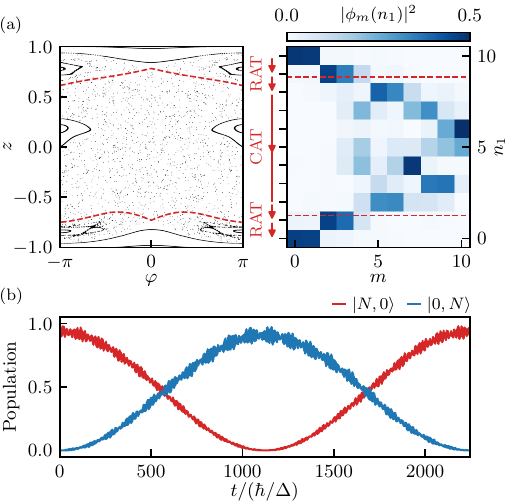}
    \caption{
    \textbf{Accelerated NOON-state preparation via resonance- and chaos-assisted tunneling.} 
    \textbf{(a)} Left: Stroboscopic Poincaré section in the $(\varphi,z)$-plane (black dots), with the partial barriers (red dashed) delimiting the regular islands near $z=\pm1$ for $\heff = 2/N = 0.2$.
    Right: Fock-space projection of the Floquet states $\ket{\phi_m}$ sorted by overlap with $\ket{N,0}$. The red dashed lines mark the $n_1$ values corresponding to the average $z$ of the partial barriers in the left sub-panel; states outside these lines are island-supported (regular), states between these lines are spread over the chaotic sea. The red arrows identifies the perturbative RAT coupling terms that enter Eq.~\eqref{eq:Veff} (each pair of small arrows amounting to $\tilde V_\deff$) and the CAT coupling across the chaotic sea (long arrow).
    \textbf{(b)} Populations of $\ket{N,0}$ (red) and $\ket{0,N}$ (blue) as a function of time, starting from $\ket{\psi_0}=\ket{N,0}$. The crossing at $t=\tau_\dNOON = 561 \, \hbar/\Delta$ realizes a NOON state with fidelity $\mathcal{F}_\dNOON = 0.947$ (Eq.~\eqref{eq:fidelity}), while the maximum $\mathcal{F}_\dNOON = 0.976$ is reached at $t=541 \,\hbar/\Delta$.
    Parameters: $\Lambda=25$, $\hbar\omega/\Delta=6.45$, $\hbar\Omega/\Delta=3.8$ and $N=10$.
    }
    \label{fig:8a}
\end{figure}

The mechanism by which the periodic drive acts on the NOON doublet is most transparently understood in the mean-field limit of Sec.~\ref{sec:selftrapping}, where the dynamics admits a classical phase-space representation. The classical limit of Eq.~\eqref{eq:H_drive}, obtained by promoting $\Omega\rightarrow\Omega\cos(\omega t)$ in Eq.~\eqref{eq:H_mf}, reads
\begin{equation}
    \label{eq:H_mf_drive}
    \tilde{H}(\tilde{t}) = \tilde{H}_0 + \tilde{H}'(\tilde{t}),
\end{equation}
with the integrable part 
\begin{equation}
    \label{eq:H_mf_0}
    \tilde{H}_0 = -\dfrac{\Lambda}{3}z^2 - \sqrt{1-z^2}\cos(\varphi),
\end{equation}
whose phase space is represented in Fig.~\ref{fig:4}~(c,d,e) for three values of the interaction parameter $\Lambda=NU/\Delta$, and with the periodic perturbation
\begin{equation}
    \label{eq:H_mf_prime}
    \tilde{H}'(\tilde{t}) = -\dfrac{2\hbar\Omega}{\Delta} \cos\left( \tilde\omega \tilde{t}\right)z,
\end{equation}
where $\tilde \omega=\hbar\omega/\Delta$ and tildes denote the dimensionless units (see Sec.~\ref{sec:selftrapping}). For moderate driving amplitudes and frequencies, the stroboscopic Poincaré section of $\tilde H(\tilde t)$~\eqref{eq:H_mf_drive} exhibits a mixed regular-chaotic structure, shown in Fig.~\ref{fig:8a}~(a) (whereas the phase space of $\tilde H_0$ is represented in Fig.~\ref{fig:4}~(e) for the same value of $\Lambda=25$). In Fig.~\ref{fig:8a}~(a), two large regular islands are inherited from the minima of $\tilde H_0$ near $z=\pm1$, hosting the self-trapped Fock states $\ket{N,0}$ and $\ket{0,N}$. These islands are surrounded by partial barriers and separated by a central chaotic sea (see below). Classically, trajectories starting inside either island remain confined to it. The two islands therefore couple only through quantum dynamical tunneling~\cite{davis_1981,hensinger_2001}, whose strength sets the doublet splitting $\Delta\varepsilon_\pm$ and hence $\tau_\dNOON$ via Eq.~\eqref{eq:NOON_time_drive}. 

Within each well of $\tilde H_0$, the motion is integrable, and conveniently parametrized by the dimensionless action-angle coordinates $(\tilde I,\tilde \theta)$, with the action variable
\begin{equation}
    \label{eq:Iz}
    \tilde I \equiv \dfrac{1}{2\pi}\oint (z+1)\,\dd\varphi \;\in [0,2],
\end{equation}
equal to the dimensionless area enclosed by a closed orbit in the $(\varphi,z)$ plane, divided by $2\pi$~\footnote{The dimensionless action $\tilde I$ relates to the physical action $I$ through $I/\hbar = \tilde I/\heff$, with $\heff = 2/N$~\cite{Note3, berezin_1975}.}. In these coordinates, $\tilde H_0$ depends on $\tilde I$ alone, and a closed orbit at action $\tilde I$ has angular frequency $\tilde \Omega_0(\tilde I) = \partial \tilde H_0/\partial \tilde I$. Semiclassical Einstein-Brillouin-Keller (EBK) quantization~\cite{wimberger_2014} yields a ladder of locally quantized quasimodes $\ket{n}$ at the actions \mbox{$\tilde I_n = (n+1/2)\heff$}, the deepest of which ($n=0$) is, to a good approximation in the self-trapping regime (Sec.~\ref{sec:selftrapping}), the Fock state $\ket{N,0}$ or $\ket{0,N}$ at the center of its respective island~\footnote{When driving the system, the centers of the self-trapping islands are generally slightly displaced from the minima ($z=\pm1$) of $\tilde H_0$. When evaluating the semiclassical RAT coupling from the center of an island to the surrounding chaotic sea, the $(\varphi,z)$ phase space (Bloch sphere) is rotated in order to place the center of the island at $z=-1 \Leftrightarrow \tilde I = 0$ (Eq.~\eqref{eq:Iz}).}.

In the extended phase space that includes time, the closed orbits of $\tilde H_0$ at action $\tilde I$ become two-dimensional tori with winding number $\alpha(\tilde I) = \tilde\omega/\tilde\Omega_0(\tilde I)$, counting  the number of modulation periods per orbital oscillation. For weak amplitudes of the non-integrable perturbation $\tilde H'$~\eqref{eq:H_mf_prime}, the Kolmogorov-Arnold-Moser (KAM) theorem~\cite{arnold_1963,wimberger_2014} states that tori with sufficiently irrational $\alpha$ are preserved as smooth deformations of the unperturbed ones. By contrast, at the discrete actions $\tilde I_\rs$ where the winding number is rational,
\begin{equation}
    \label{eq:resonance_condition}
    \alpha(\tilde I_\rs) = r/s \quad \Leftrightarrow \quad r \tilde\Omega_0(\tilde I_\rs) = s \tilde \omega,
\end{equation}
with $r,s\in \mathbb{N}^*$, the corresponding torus is
generically destroyed by the Poincaré-Birkhoff mechanism~\cite{wimberger_2014}, breaking into a chain of $r$ elliptic sub-islands separated by $r$ hyperbolic points: a nonlinear $\rs$-resonance.

Beyond the outermost resonance chain of each island, neighboring resonances overlap and the surrounding KAM tori are destroyed (Chirikov criterion~\cite{chirikov_1979}), giving rise to the connected chaotic sea visible between the two polar islands in Fig.~\ref{fig:8a}~(a). The last invariant structures separating the regular core of each island from this chaotic sea are partial barriers~\cite{bohigas_1993,schlagheck_2011}. Although partial barriers no longer prevent classical trajectories from crossing them, they remain effective quantum frontiers when $\heff$ is larger than the phase-space flux periodically exchanged through the barrier. This regular-chaotic dichotomy is directly visible across the two sub-panels of Fig.~\ref{fig:8a}~(a), which share a common vertical axis (population $n_1=N(z+1)/2$): on the classical side (left), the partial barriers appear as the red dashed curves bounding each island; on the quantum side (right), they appear as the boundary between symmetry-paired Floquet doublets with narrow Fock support on $\ket{n_1,N-n_1}$ and $\ket{N-n_1,n_1}$ (left-most columns) and Floquet states broadly distributed over the intermediate Fock range, corresponding to ``chaotic'' states delocalized over the chaotic sea. Denoting by $\tilde I_c$ the action of the partial barrier, the regular core of an island consists of EBK quasimodes $\ket{n}$ with $\tilde I_n < \tilde I_c$. Beyond $\tilde I_c$, the dynamics is governed by the chaotic Floquet states of the central sea, forming a block of $N_c$ levels (Fig.~\ref{fig:8a}~(a)) approximately uniformly distributed over the Floquet quasienergy zone $[0,\hbar \omega)$. 

In the self-trapping regime that isolates $\ket{N,0}$ and $\ket{0,N}$ from the rest of the spectrum, $\Delta\varepsilon_\pm$ is generally small. As we detail below, it can be strongly enhanced by the intermediate phase-space structures, with RAT from the center of an island to the chaotic sea, and CAT through the latter.

\subsubsection{Resonance- and chaos-assisted tunneling}
\label{subsub:ratcat}
In the vicinity of a nonlinear $\rs$-resonance chain inside the regular island ($\tilde I < \tilde I_c$), the driven Hamiltonian~\eqref{eq:H_mf_drive} admits the effective description~\cite{eltschka_2005, schlagheck_2011}
\begin{equation}
    \label{eq:H_mf_eff}
    \tilde H_\deff^{(\rs)}(\tilde I,\tilde \theta) = \dfrac{(\tilde I-\tilde I_\rs)^2}{2\tilde m_\rs} + 2 \tilde V_\rs\text{cos} (r\tilde \theta),
\end{equation}
parametrized by the resonance action $\tilde I_\rs$, the effective mass $\tilde m_\rs$ and the resonance coupling strength $\tilde V_\rs$~\footnote{$\tilde I_\rs$, $\tilde m_\rs$ and $\tilde V_\rs$ express the corresponding physical quantities $I_\rs$, $m_\rs$ and $V_\rs$ in units of $\hbar/\heff=N\hbar/2$, $N\hbar^2/(2\Delta)$ and $N\Delta/2$ respectively.}, all directly obtained from the Poincaré section of $\tilde H(t)$~\eqref{eq:H_mf_drive}~\cite{eltschka_2005,schlagheck_2011}. Quantizing Eq.~\eqref{eq:H_mf_eff}, the $\rs$-resonance perturbation couples EBK quasimodes $\ket{n}$ and $\ket{n+r}$, whose reduced action differs by $r\heff$, through the action-dependent matrix elements~\cite{schlagheck_2011}
\begin{equation}
    \label{eq:V_rs_elements}
    \langle n+r | \hat{\tilde H}_\deff^{(\rs)} | n \rangle \equiv \tilde V_\rs^{(n+r)} = \tilde V_\rs \left(\dfrac{\heff}{\tilde I_{r:s}}\right)^{r/2}\sqrt{\dfrac{(n+r)!}{n!}}.
\end{equation}
Iterating this perturbative coupling from the deepest quasimode $\ket{n=0}\simeq\ket{N,0}$ (or $\ket{0,N}$) up to the edge of the regular island --- a perturbative process depicted in Fig.~\ref{fig:8a}~(a) with the short vertical red arrows and detailed in Ref.~\cite{schlagheck_2011} --- yields the effective matrix element coupling the NOON Fock states to the chaotic sea:
\begin{equation}
    \label{eq:Veff}
    \tilde{V}_\deff = \tilde{V}_{r{:}s}^{[(k_c+1)r]} \prod_{k=1}^{k_c} \dfrac{\tilde{V}_{r{:}s}^{(kr)}}{\tilde{E}_0-\tilde{E}_{kr}+ks\heff\tilde\omega},
\end{equation}
where $k_c = \lfloor (\tilde I_c/\heff - 1/2)/r\rfloor$ is the largest integer such that the quasimode $\ket{k_cr}$ at action $\tilde I_{k_cr}=(k_cr+1/2)\heff$ lies inside the partial barrier, and where $\tilde E_n = E_n/(N\Delta/2)$ are the unperturbed energies of the  quasimodes $\ket{n}$, obtained from exact diagonalization of $\hH_0$ at $\Omega=0$~\footnote{Because $\hH_0$ exhibits quasi-degenerate doublets in the self-trapping regime (see Fig.~\ref{fig:4}~(a)), the $n$-th EBK quasimode of one island corresponds to the $(2n)$-th eigenstate of $\hH_0$. $E_{kr}$ in Eq.~\eqref{eq:Veff} is therefore the $(2kr)$-th eigenvalue of $\hH_0$.}.

The effective coupling $\tilde V_\deff$ obtained from RAT connects each NOON Fock state to the chaotic sea at the chaos edge $\tilde I_c$. Beyond this edge, the dynamics is no longer described by the resonance Hamiltonian~\eqref{eq:H_mf_eff} but by the chaotic Floquet states of the central sea. By the generalized parity $\hPi$ of the driven system, the two islands play interchangeable roles: each chaotic state has, by ergodicity, generic Husimi support at the chaos edge of both islands, and thus couples to both $\ket{N,0}$ and $\ket{0,N}$ through $\tilde V_\deff$. Amplitude leaking from $\ket{N,0}$ into the chaotic block via $\tilde V_\deff$ propagates over the $N_c$ chaotic levels (as depicted by the long vertical red arrows in Fig.~\ref{fig:8a}~(a)) and couples back into $\ket{0,N}$ through a second factor of $\tilde V_\deff$. At second order in $\tilde V_\deff$, this two-step process produces a splitting between the quasienergies of the symmetric and antisymmetric NOON Floquet states $\ket{\phi_\pm}$. Under the random-matrix-theoretic (Gaussian orthogonal ensemble) assumption for the chaotic block~\cite{tomsovic_1994,schlagheck_2011}, the typical splitting evaluates to~\cite{eltschka_2005,schlagheck_2011}
\begin{equation}
    \label{eq:DE_RATCAT}
    \Delta \varepsilon_\pm \simeq \dfrac{2\pi}{\hbar\omega} \left( \dfrac{N\Delta}{2}\tilde{V}_\deff\right)^2.
\end{equation}
Finally, the RAT-CAT estimate for the NOON-creation time follows from Eq.~\eqref{eq:NOON_time_drive}.

\subsubsection{NOON state preparation from RAT and CAT}
\label{subsub:ratcat_results}
\begin{figure}[!t]
    \centering
    \includegraphics[scale=1]{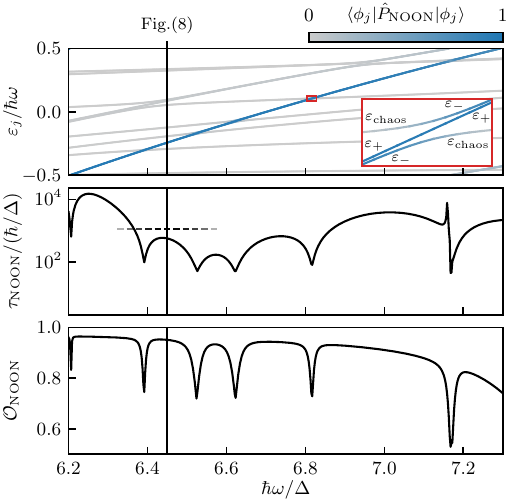}
    \caption{
    \textbf{Spectral and dynamical signatures of the NOON doublet across modulation frequency.}
    Top: Floquet quasienergies $\varepsilon_j$ as a function of the modulation frequency $\hbar\omega/\Delta$, color-coded from gray to blue by overlap with the NOON subspace, $\langle\phi_j|\hP_\dNOON|\phi_j\rangle$ (Eq.~\eqref{eq:NOON_projectors}), identifying the Floquet doublet $\ket{\phi_\pm}$ (first two columns of the Fock projection in Fig.~\ref{fig:8a}~(a)) supporting this subspace, except for avoided crossings (one such crossing is magnified in the inset).
    Middle: NOON-creation time $\tau_\dNOON$ extracted from the splitting of the NOON-supporting doublet as a function of the modulation frequency. The horizontal dashed line at $\tau_\dNOON = 1105 \,\hbar/\Delta$ is the analytical resonance-assisted tunneling estimate at $\hbar\omega/\Delta = 6.45$ (operating point of Fig.~\ref{fig:8a}).
    Bottom: Floquet-NOON overlap $\mathcal{O}_\dNOON$ (Eq.~\eqref{eq:overlap}) as a function of $\hbar\omega/\Delta$; its drops at avoided crossings bound the attainable NOON-state fidelity.
    Parameters: $\Lambda=25$, $\hbar\Omega/\Delta=3.8$ and $N=10$.
    }
    \label{fig:8b}
\end{figure}

Figure~\ref{fig:8b} summarizes the phenomenology of the RAT-CAT acceleration as the modulation angular frequency $\omega$ is varied, at a fixed amplitude of modulation $\hbar\Omega = 3.8\,\Delta$. As established in Sec.~\ref{subsub:driven_ratcat}, this amplitude sets the degree of non-integrability: it must be large enough to develop the connected chaotic sea and the nonlinear resonances that mediate the tunneling enhancement, yet small enough to preserve the regular islands anchoring $\ket{N,0}$ and $\ket{0,N}$. We therefore use a moderate amplitude that balances this enhancement against the hybridization of the NOON doublet with the chaotic sea quantified below, producing the mixed phase space of Fig.~\ref{fig:8a}~(a). The Floquet quasienergies $\varepsilon_j$ (Fig.~\ref{fig:8b}, top) are color-coded by their projection on the NOON subspace $\langle\phi_j|\hP_\dNOON|\phi_j\rangle$, with
\begin{equation}
    \label{eq:NOON_projectors}
    \hP_\dNOON = |N,0\rangle\langle N,0|+|0,N\rangle\langle 0,N|.
\end{equation}
The Floquet doublet $\ket{\phi_\pm}$ retains dominant NOON support across the scanned range, with the exception of narrow avoided crossings where a chaotic Floquet state of the appropriate $\hPi$-parity becomes locally resonant with the doublet (inset of Fig.~\ref{fig:8b}, top). At each crossing, the level splitting $\Delta\varepsilon_\pm$ is strongly enhanced~\cite{tomsovic_1994,schlagheck_2011,arnal_2020,martinez_2021}, producing the pronounced dips in the NOON time of the middle panel: $\tau_\dNOON$ (Eq.~\eqref{eq:NOON_time_drive}) ranges from $\sim10^4\,\hbar/\Delta$ off-resonance down to $\sim 10^2\,\hbar/\Delta$ at the deepest crossings, many orders of magnitude below the static value $\tau_\dNOON(\Omega=0) = 2.93\cdot10^9\,\hbar/\Delta$ obtained in Sec.~\ref{sec:selftrapping} for the same $\Lambda$ and $N$. However, whether such a fast splitting realizes a clean NOON dynamics depends on how completely the Floquet doublet $\{\ket{\phi_\pm}\}$ spans the NOON subspace~\cite{vanhaele_2021,dupont_2023_a}. We quantify this by the Floquet-NOON overlap
\begin{equation}
\label{eq:overlap}
    \mathcal{O}_{\dNOON} = \dfrac{1}{2}\Tr\left(\hP_\dNOON \hP_\pm\right),
\end{equation}
with $\hP_\pm = |\phi_+\rangle\langle \phi_+|+|\phi_-\rangle\langle \phi_-|$. This overlap equals 1 when the two subspaces coincide and drops at the avoided crossings as one of the NOON-supporting states hybridizes with a chaotic state~\cite{dupont_2023_b} (Fig.~\ref{fig:8b}, bottom). Off-resonance, $\mathcal{O}_\dNOON \gtrsim 0.95$ is maintained alongside $\tau_\dNOON\simeq10^3\,\hbar/\Delta$, six orders of magnitude faster than the static prediction, with negligible fidelity loss to the chaotic block. The proximity of $\mathcal{O}_\dNOON$ to unity at $\hbar \Omega = 3.8 \,\Delta$ confirms that this amplitude lies in the favorable part of the tradeoff: large enough for fast tunneling, small enough to preserve the NOON doublet.

We now apply the semiclassical RAT-CAT formulas of Eqs.~\eqref{eq:Veff} and~\eqref{eq:DE_RATCAT} to the operating point $\hbar\omega/\Delta = 6.45$ of Fig.~\ref{fig:8a} (away from avoided crossings). The Poincaré section of Fig.~\ref{fig:8a}~(a)-left features a prominent 1:2 resonance chain near $|z|\approx0.8$ in each island, between its central NOON Fock state and the partial barrier. From the exterior and interior frontiers of this chain~\cite{eltschka_2005,schlagheck_2011}, we extract its average action $\tilde I_{1:2} = 0.214$ and coupling strength $\tilde V_{1:2} = 0.0487$ (see Eq.~\eqref{eq:H_mf_eff}). The outer separatrix of this same chain delimits the regular island: the partial barrier accordingly lies just beyond the chain, at action $\tilde I_c = 0.304$, giving $k_c=1$ for $\heff=0.2$ and $r=1$: the center of the island is coupled to the chaotic sea via a two-step ladder of quasimodes, with $\ket{0}\rightarrow\ket{1}$ inside the partial barrier, then $\ket{1}\rightarrow\ket{2}$ across the partial barrier into the chaotic sea (a process depicted in Fig.~\ref{fig:8a}~(a) by the short vertical arrows between the panels). In combination with CAT across the chaotic sea, the resulting RAT coupling yields the NOON-creation time $\tau_\dNOON = 1105 \,\hbar/\Delta$ (Eqs.~\eqref{eq:Veff}, \eqref{eq:DE_RATCAT} and~\eqref{eq:NOON_time_drive}), and is to be compared with the exact value $\tau_\dNOON^{\text{num}}=561 \,\hbar/\Delta$ extracted from the diagonalization of the Floquet operator at this operating point. The two agree within a factor of two, on a quantity that varies by more than three orders of magnitude across the scanned frequency range. As already noted in Ref.~\cite{vanhaele_2021}, this level of agreement is satisfactory and confirms that the 1:2 resonance is here the dominant perturbative pathway connecting the NOON doublet to the chaotic sea. Clearly, the semiclassical theory in its present form is not quantitatively predictive, in the sense that a determination of $\tau_\mathrm{NOON}$ that is sufficiently precise for the purpose of experimental implementations requires the full Floquet diagonalization. Possible routes for the improvement of the semiclassical prediction include a more refined modeling of the 1:2 resonance coupling, beyond the simple pendulum description \cite{schlagheck_2011}, as well as a more detailed treatment of the chaotic block beyond its random-matrix average.

Figure~\ref{fig:8a}~(b) demonstrates the concrete realization of accelerated NOON preparation through RAT and CAT: starting from $\ket{\psi_0}=\ket{N,0}$ the populations of $\ket{N,0}$ and $\ket{0,N}$ exchange completely under the driven Hamiltonian~\eqref{eq:H_drive}. At $\tau_\dNOON = 561 \,\hbar/\Delta$, a NOON state is realized with fidelity $\mathcal{F}_\dNOON = 0.947$. Additional fast oscillations, resulting from tiny couplings with other Floquet states, have the effect that the maximum of fidelity $\mathcal{F}_\dNOON=0.976$ is reached slightly earlier, at $t=541 \,\hbar/\Delta$. The fidelity ceiling of $0.95-0.98$, set by $\mathcal{O}_\dNOON$ for these parameters, is consistent with the bound discussed above.

\section{Conclusion}

In this work, we demonstrated that microscopic vortex NOON states can be efficiently prepared in a weakly anisotropic two-dimensional trap and exploited as quantum-enhanced probes for rotation sensing. In the self-trapping regime, interactions isolate the NOON manifold and provide both protection against leakage and Heisenberg-limited sensitivity. To overcome the exponentially slow collective tunneling, we introduced two complementary acceleration strategies. For small particle numbers, geodesic counterdiabatic driving enables the deterministic preparation of high-fidelity NOON states on experimentally relevant timescales. For larger particle numbers, resonance- and chaos-assisted tunneling provide an alternative route in the near-semiclassical regime, leading to a strong enhancement of the tunneling dynamics --- a mechanism in which the chaos-assisted component has already been observed with Bose-Einstein condensates in a driven optical lattice~\cite{arnal_2020}. The preparation protocols developed here, which assume loading into the p-state manifold~\cite{isacsson_2005, Liu2006,Muller2007, Wirth_2010,Wang_2021}, also provide natural starting points for further optimization using quantum optimal control techniques~\cite{ansel_2024,lapert_2012,dupont_2021,dupont_2024}, with the GCD or RAT-CAT protocols serving as initial guesses for the optimization procedure. Such an approach could further reduce preparation times, thereby increasing robustness against decoherence and unwanted background rotations while extending the accessible particle-number regime.

The interferometric protocol proposed here provides a direct operational use of vortex NOON states for rotation sensing. By preparing a coherent superposition of opposite circulation states, allowing a phase to accumulate under an external rotation, and recombining the two components, the protocol leads to a measurable population imbalance. Combined with the enhanced phase sensitivity of NOON states, this provides a realistic pathway toward Heisenberg-limited rotation sensing with interacting ultracold atoms. Further improvement of the sensitivity can be achieved by creating such microscopic vortex NOON states within each of the $L$ wells of an optical lattice~\cite{stolzenberg_2025}, controlling $\Delta$ through the relative intensity of the laser-beam pairs, and thereby allowing one to probe the rotation $\Omega$ through $\nu=L$ parallel realizations of the NOON sensor (see Eq.~\eqref{eq:QCRB}). This further reduces the uncertainty on the estimation of $\Omega$ by a factor $\sqrt{L}$~\cite{pezze_2018}.

Beyond the specific realization considered here, the preparation and metrological protocols developed in this work are applicable to any effective two-mode platform supporting states with opposite circulation. This includes, for instance, ring-shaped Bose–Einstein condensates threaded by synthetic gauge fields~\cite{aghamalyan2015coherent,Nicolau_2020,Pradhan_2024, Carmona-Lopez2026} and $\pi$-flux plaquettes~\cite{DiLiberto_pi_flux}, where coherent superpositions of counter-propagating persistent-current states can also be exploited for quantum-enhanced rotation sensing. The present toolbox has thus a wide potential utility in the context of quantum sensing with trapped bosonic atoms.

\begin{acknowledgments}
This work was supported by the ERC Grant LATIS, the EOS Project CHEQS (EOS 40007526) and the ANR PEPR Grants QUTISYM ANR-23-PETQ-0002 and Dyn1D ANR-23-PETQ-0001. S.D. acknowledges funding from the European Union’s Horizon Europe
Framework Programme (EIC Pathfinder Challenge project Veriqub) under Grant Agreement
No. 101114899.
\end{acknowledgments}

\appendix

\section{Angular momentum in an anisotropic 2D harmonic potential}
\label{app:Lz_2D}

We detail here the action of the angular momentum operator $\hL_z^{(\ob)} = \hx\hp_y - \hy\hp_x$ on the vortex states $\ket{\psi_\pm}$~\eqref{eq:psi_pm}, justifying both the eigenvalue $\pm R\hbar$ quoted in Sec.~\ref{sec:bh_model} and the sense in which $\ket{\psi_\pm}$ are approximate eigenstates of $\hL_z^{(\ob)}$.

Expressed through the Cartesian ladder operators of Eq.~\eqref{eq:H0_1body_ladder}, the angular momentum reads
\begin{equation}
\label{eq:Lz_ladder}
\begin{split}
    \hL_z^{(\ob)} &= \di\hbar\,\frac{\wx+\wy}{2\sqrt{\wx\wy}}\left(\ha_y^\dagger\ha_x - \ha_x^\dagger\ha_y\right)\\
    &+ \di\hbar\,\frac{\wy-\wx}{2\sqrt{\wx\wy}}\left(\ha_x^\dagger\ha_y^\dagger - \ha_x\ha_y\right),
\end{split}
\end{equation}
with the same structure as the rotation term of Eq.~\eqref{eq:H0_1body_ladder}: a first contribution that preserves the total excitation number $n_x+n_y$, and a second, proportional to the anisotropy $(\wy-\wx)$, that changes it by $\pm 2$. Acting on the p-orbital states $\ket{\dpx}\equiv\ket{\phi_{10}}$ and \mbox{$\ket{\dpy}\equiv\ket{\phi_{01}}$}, the first term couples them within the manifold, $\ha_y^\dagger\ha_x\ket{\phi_{10}} = \ket{\phi_{01}}$ and $\ha_x^\dagger\ha_y\ket{\phi_{01}} = \ket{\phi_{10}}$, while the second promotes them to the third excited manifold, $\ha_x^\dagger\ha_y^\dagger\ket{\phi_{10}} = \sqrt{2}\,\ket{\phi_{21}}$ and $\ha_x^\dagger\ha_y^\dagger\ket{\phi_{01}} = \sqrt{2}\,\ket{\phi_{12}}$ (see Fig.~\ref{fig:2}). Applying Eq.~\eqref{eq:Lz_ladder} to the vortex states~\eqref{eq:psi_pm} then gives
\begin{equation}
\label{eq:Lz_psi}
\begin{split}
    \hL_z^{(\ob)}\ket{\psi_\pm} = &\pm\hbar\,\frac{\wbar}{\sqrt{\wx\wy}}\ket{\psi_\pm}\\
    &\mp \hbar\,\frac{\wy-\wx}{\sqrt{2\wx\wy}}\,\frac{\ket{\phi_{12}}\mp\di\ket{\phi_{21}}}{\sqrt{2}}.
\end{split}
\end{equation}
The first term identifies $\ket{\psi_\pm}$ as quasi-eigenstates of $\hL_z^{(\ob)}$ with eigenvalue $\pm R\hbar$, where $R\equiv\wbar/\sqrt{\wx\wy}$ and $\wbar=(\wx+\wy)/2$; in the weak-anisotropy regime~\eqref{eq:weak_anisotropy}, $\wbar\simeq\sqrt{\wx\wy}$ and hence $R\simeq 1$. The second term is the leakage of $\ket{\psi_\pm}$ into the third excited manifold, suppressed by the small ratio $(\wy-\wx)/\sqrt{2\wx\wy}$. It is the one-body counterpart of the inter-manifold coupling neglected in Eq.~\eqref{eq:H0_1body_ladder} under condition~\eqref{eq:weak_anisotropy}, and it is in this sense that $\ket{\psi_\pm}$ are eigenstates of $\hL_z^{(\ob)}$ to a good approximation, with corrections of order $(\wy-\wx)/\wbar \ll 1$.

\section{Derivation of the Bose-Hubbard model}
\label{app:bose_hubbard_model}

We detail the passage from the trapped rotating Bose gas described by the Hamiltonian~\eqref{eq:H0_1body} to the two-mode Bose-Hubbard Hamiltonian~\eqref{eq:H_BH}, and give the interaction energy $U$ in terms of the trap frequencies and the $s$-wave scattering length $a_s$.

We consider a strong axial confinement along the third dimension $z$ of the trap: $\omega_z \gg \omega_{x},\omega_y$, which freezes the $z$ motion in the oscillator ground state \mbox{$\phi_0(z) = (m\omega_z/\pi\hbar)^{1/4} \exp\{-m\omega_z z^2/2\hbar\}$}. Writing \mbox{$\hpsi(\bm{r}) = \hpsi(x,y)\,\phi_0(z)$} in the contact
interaction integrals (see Eq.~\eqref{eq:H_BH_int} and Refs.~\cite{dalfovo_1999,li_2016}), with $g = 4\pi\hbar^2 a_s/m$, the axial integral factorizes and yields the effective two-dimensional interaction parameter
\begin{equation}
    \label{eq:g2D}
    g_{2\dD} = g\!\int_{-\infty}^\infty\!\dd z\,|\phi_0(z)|^4 = 2a_s\sqrt{\frac{2\pi\hbar^3\omega_z}{m}}.
\end{equation}

Along $x$ and $y$ in the p-orbital manifold, the bosonic field reads 
\begin{equation}
\label{eq:bosonic_field}
\begin{split}
    \hpsi(x,y) &= \dpx(x,y)\hb_x + \dpy(x,y)\hb_y\\
    &= \psi_+(x,y)\hb_+ + \psi_-(x,y)\hb_-,
\end{split}
\end{equation}
with 
\begin{align}
    \label{eq:pxpy}
        \dpx(x,y) \equiv \phi_{10}(x,y) = \sqrt{\dfrac{2}{\pi}} \dfrac{m}{\hbar} \left(\omega_x^3\omega_y\right)^{1/4}\\
        \notag\times x\,\exp\left\{-\dfrac{m}{2\hbar}\left(\omega_x x^2 + \omega_y y^2\right)\right\},\\
       \label{eq:pxpy2} \dpy(x,y) \equiv \phi_{01}(x,y) = \sqrt{\dfrac{2}{\pi}} \dfrac{m}{\hbar} \left(\omega_x\omega_y^3\right)^{1/4}\\
        \notag\times y\,\exp\left\{-\dfrac{m}{2\hbar}\left(\omega_x x^2 + \omega_y y^2\right)\right\},
\end{align}
and the vortices \mbox{$\psi_\pm(x,y) \equiv (\dpx(x,y) \pm \di\dpy(x,y))/\sqrt{2}$} (Eq.~\eqref{eq:psi_pm}) giving $\hb_\pm^\dagger = (\hb_x^\dagger \pm \di \hb_y^\dagger)/\sqrt{2}$. Projecting $\hH_\ob$~\eqref{eq:H0_1body_ladder} onto this manifold directly yields Eq.~\eqref{eq:H_BH_0}, with \mbox{$(E_{10}+E_{01})/2 = 2\hbar\wbar$}, $E_{10}-E_{01}=-\Delta$,
\mbox{$\hn_x-\hn_y = \hb_+^\dagger\hb_- + \hb_-^\dagger\hb_+$} and
\begin{equation}
    \hL_z = \di R\hbar(\hb_y^\dagger\hb_x - \hb_x^\dagger\hb_y) = R\hbar(\hn_+ - \hn_-).
\end{equation}

The interaction Hamiltonian~\eqref{eq:H_BH_int} is derived from the two-mode bosonic field~\eqref{eq:bosonic_field} as
\begin{equation}
    \hH_\dint = \dfrac{g_{2\dD}}{2} \iint_{-\infty}^\infty \dd x \,\dd y \, \hpsi^{\dagger2}(x,y)\hpsi^2(x,y).
\end{equation}
Expanding the fields in the $(x,y)$-basis, only even combinations of the odd wavefunctions $\dpx(x,y)$ and $\dpy(x,y)$ survive the integral (see Eqs.~\eqref{eq:pxpy} and ~\eqref{eq:pxpy2}). As these functions are real, only two terms remain:
\begin{equation}
\begin{split}
    \iint_{-\infty}^\infty \dd x\,\dd y\,\dpx^4(x,y) &= \iint_{-\infty}^\infty \dd x\,\dd y\,\dpy^4(x,y)\\ &= \dfrac{3}{8\pi}\dfrac{m}{\hbar}\sqrt{\omega_x\omega_y},\\
    \iint_{-\infty}^\infty \dd x\,\dd y\,\dpx^2(x,y)\dpy^2(x,y) &= \dfrac{1}{8\pi}\dfrac{m}{\hbar}\sqrt{\omega_x\omega_y},
\end{split}
\end{equation}
which finally gives Eq.~\eqref{eq:H_BH_int}:
\begin{equation}
\begin{split}
    \hH_\dint = \,&\dfrac{U}{2}(\hb_x^{\dagger2}\hb_x^2+\hb_y^{\dagger2}\hb_y^2) + \dfrac{U}{6}(\hb_x^{\dagger2}\hb_y^2 + \hb_y^{\dagger2}\hb_x^2)\\
    &+ \dfrac{2U}{3}(\hn_x\hn_y)\\
    =&-\dfrac{U}{3} \left( \hb_+^{\dagger2}\hb_+^2 + \hb_-^{\dagger2}\hb_-^2 \right) + \dfrac{2}{3}U\hN(\hN-1),
\end{split}
\end{equation}
where the interactions become non-trivially diagonal in the vortex basis~\cite{li_2016}, and with
\begin{equation}
    \label{eq:U_parameter}
    U = \dfrac{3a_s}{2}\sqrt{\dfrac{m\hbar}{2\pi}}\sqrt{\omega_x\omega_y\omega_z}.
\end{equation}
Importantly, Eq.~\eqref{eq:U_parameter} shows that $U$ scales as $\sqrt{\omega_x\omega_y}$ (at fixed $\omega_z$), which is the same in-plane scale appearing in the weak-anisotropy condition~\eqref{eq:weak_anisotropy}. 
This condition thus plays a double role: first, it suppresses leakage out of the p-orbital manifold (see Sec.~\ref{sec:bh_model} and Appendix ~\ref{app:Lz_2D}), which validates the two-mode description underlying Eq.~\eqref{eq:H_BH}. Second, it makes $\Delta$ small relative to the in-plane scale $\sqrt{\omega_x\omega_y}$ that also sets $U$, working toward the self-trapping regime $NU/\Delta \gg 1$ (see Sec.~\ref{sec:selftrapping}). Reaching this regime additionally requires sufficient interaction strength (scattering length $a_s$ and out-of-plane confinement $\omega_z$) and particle number $N$.

\section{Experimental parameters}
\label{app:exp_param}

The question that now arises is whether the typical frequency values considered in this work, and in particular the hierarchy (\ref{eq:C5_hierarchy}), are plausible within a realistic experimental setting. In the article \cite{Hadzibabic_2006}, the authors investigate a bosonic gas of rubidium~87 confined in a quasi two-dimensional harmonic potential, realized using a one-dimensional optical lattice. The trapping frequencies associated with the three spatial directions are
\begin{align}
&\omega_x/2\pi = 11\,\mathrm{Hz}, \\
&\omega_y/2\pi = 130\,\mathrm{Hz}, \\
&\omega_z/2\pi = 3600\,\mathrm{Hz},
\end{align}
which corresponds to a very strong confinement along the $z$ axis, ensuring the quasi two-dimensional character of the system, and to a pronounced anisotropy in the transverse $(x,y)$ plane. In this configuration, the total number of rubidium~87 atoms trapped in each optical well is of the order of $N \sim 10^{5}$. 

For the atomic species $^{87}$Rb, the contact interaction can be rewritten in terms of the $s$-wave scattering length, $a_s \simeq 5.29\cdot10^{-9}\,\mathrm{m}$, and the atomic mass $m = 1.44\cdot10^{-25}\,\mathrm{kg}$ \cite{Jaksch1998}, leading to an effective interaction strength
\begin{equation}
U/\hbar \simeq 4.19\,\mathrm{rad/s}.
\end{equation}
The anisotropy associated with the trap is given by $ \Delta = \hbar(\omega_y - \omega_x)$, which leads to the numerical value
\begin{equation}\label{eq:estim_delta}
\Delta/\hbar \simeq 7.48\cdot10^{2}\,\mathrm{rad/s}.
\end{equation}
At first glance, this anisotropy appears to be much larger than the two-body interaction scale. However, it should be recalled that the self-trapping regime is a collective phenomenon, whose relevant criterion involves the total number of particles. The appropriate quantity to consider is therefore the ratio $NU/\Delta$. Under the experimental conditions of \cite{Hadzibabic_2006}, one finds $NU/\Delta \approx 560$,
indicating that the self-trapping regime is largely satisfied in this experimental setup. As an order-of-magnitude estimate, a ratio $NU/\Delta \approx 10$ could already be achieved for a particle number of the order of $N \sim 10^{3}$. The self-trapping regime may also be reached by reducing the anisotropy $\Delta$, for instance by choosing a trapping frequency $\omega_x$ closer to $\omega_y$.

Given estimation (\ref{eq:estim_delta}), one can determine the order of magnitude for different kind of external rotation. For example, the angular frequency associated with Earth's rotation is given by $ \omega_\mathrm{\oplus}=7\cdot 10^{-5}\,\mathrm{rad/s}$, leading to a perturbation amplitude of
\begin{equation}
    \hbar\Omega_\oplus/\Delta \simeq 9.3\cdot10^{-8}.
\end{equation}
Using Eq.~(\ref{eq:QCRB}), one can therefore estimate the number of measurements $\nu$ needed to achieve a sensitivity that would allow to detect Earth's rotation. For $N=10$ particles and an interrogation time of $T=10^4\,\hbar/\Delta$, a minimal number of measurements $\nu_\mathrm{min} \simeq 2.9\cdot10^{3}$ would be needed. Such a value is compatible with optical-lattice architectures \cite{Tao2024}, where hundreds of identical NOON-state interferometers could in principle be operated in parallel.

For the previously established values, the frequency hierarchy (\ref{eq:C5_hierarchy}) can also be fixed for an arbitrary GCD protocol duration $T = 100\,\hbar/\Delta$ and for $N=10$. In this case, the different characteristic frequencies of the system read
\begin{align}
    &\omega_\mathrm{GCD} \sim 1/T = \Delta/(100\hbar) \approx 7.5\,\mathrm{rad/s}, \\
    &\omega_\mathrm{MB} \sim 2U(N-1)/(3\hbar) \approx 25\,\mathrm{rad/s}, \\
    &\bar{\omega} \approx 2\pi\cdot440\,\mathrm{rad/s}.
\end{align}
These order-of-magnitude estimates confirm that the parameter regime considered here is compatible with experimentally accessible conditions, while allowing for a clear separation between the different timescales of the problem. The hierarchy (\ref{eq:C5_hierarchy}) therefore provides a consistent framework for analyzing the dynamics induced by the GCD protocol.

\section{Perturbation theory of the NOON coupling in the self-trapping regime}
\label{app:perturbation_theory}

To define the reduced Hamiltonian~\eqref{eq:hred} governing the Hilbert subspace $\{\ket{N,0},\ket{0,N}\}$ in the self-trapping regime, we determine how the coupling of these states to the rest of the Hilbert space lifts the degeneracy between them. Setting $\Omega=0$, we split the Bose-Hubbard Hamiltonian~\eqref{eq:BHMomega} as $\hH = \hH_\dint + \hV$, where
\begin{equation}
    \hH_\dint = -\dfrac{U}{3}\left(\hb_+^{\dagger2}\hb_+^2 + \hb_-^{\dagger2}\hb_-^2\right)
\end{equation}
is the interaction term, diagonal in the Fock basis with eigenvalues
\begin{equation}
    \begin{split}
        E_{n_+,n_-} &\equiv \bra{n_+,n_-}\hH_\dint\ket{n_+,n_-}\\ &= -\dfrac{U}{3}\left[2n_+(n_+-N)+N(N-1)\right],
    \end{split}
\end{equation}
with $n_- = N-n_+$, and
\begin{equation}
    \hat{V} = -\dfrac{\Delta}{2} \left( \hb_+^\dagger\hb_- + \hb_-^\dagger \hb_+ \right)
\end{equation}
is the anisotropy-induced coupling, treated as a perturbation in the self-trapping regime $NU/\Delta \gg 1$. The interaction $\hH_\dint$ is symmetric under $+\leftrightarrow-$, so $\ket{N,0}$ and $\ket{0,N}$ share the same energy $E_{N,0}=E_{0,N}$. The perturbation $\hV$ couples them through the rest of the Hilbert space and lifts the degeneracy, enabling collective tunneling. Since $\hV$ transfers one boson at a time between the two modes, and all $N$ bosons must change mode, the leading contribution connecting $\ket{N,0}$ and $\ket{0,N}$ is of order $N$, generated along the unique path
\begin{equation}
    \ket{0,N} \to \ket{1,N-1} \to \ldots  \to \ket{N-1,1} \to \ket{N,0}.
\end{equation}
The off-diagonal element of the reduced Hamiltonian $\hat{H}_\mathrm{red}$ connecting the two states is then
\begin{equation}
\label{eq:H_red_pert}
\begin{split}
    \bra{0,N} &\hH_\dred \ket{N,0} = \\
    &\frac{\langle 0,N\vert \hat{V}\vert 1,N-1\rangle \ldots \langle N-1,1\vert\hat{V}\vert N,0\rangle}{(E_{N,0}-E_{1,N-1})\ldots (E_{N,0}-E_{N-1,1})}.
\end{split}
\end{equation}
For $U>0$, each factor in the denominator of Eq.~\eqref{eq:H_red_pert} is thus negative, with \mbox{$E_{N,0}-E_{k,N-k} = -2Uk(N-k)/3<0$}. The product of these $N-1$ terms therefore carries a factor $(-1)^{N-1}$, while the numerator, built from $N$ hopping amplitudes each proportional to $-\Delta/2$, carries $(-1)^N$. They combine to give a matrix element
\begin{equation}
    \bra{0,N}\hH_\dred\ket{N,0} = -\frac{N(\Delta/2)^N}{(N-1)!(2U/3)^{N-1}},
\end{equation}
that is negative for all $N$, irrespective of parity. With the convention of Eq.~\eqref{eq:hred}, $\hH_\dred$ carries $-\mathcal{J}$ on its off-diagonal, so
\begin{equation}
    \mathcal{J}=\frac{N(\Delta/2)^N}{(N-1)!(2U/3)^{N-1}}>0.
\end{equation}
This sign is fixed by the attractive nature of the interaction: it places the symmetric NOON state $(\ket{N,0}+\ket{0,N})/\sqrt{2}$ as the ground state, in agreement with the exact diagonalization of~\eqref{eq:BHMomega} for both even and odd $N$. Note that a repulsive interaction would reverse the sign of the denominators, making the sign of $\mathcal{J}$ alternate with $N$ and exchanging which NOON superposition is the ground state. As each order corresponds to one "back-and-forth" process of a boson, the natural structure of higher orders in perturbation theory is of the form
\begin{equation}
    \mathcal{J} = \frac{N(\Delta /2)^N}{(N-1)!(2U/3)^{N-1}}\Big[1+ c_1\Big(\frac{\Delta}{U}\Big)^2 +c_2\Big(\frac{\Delta}{U}\Big)^4 + \ldots  \Big],
\end{equation}
where the $c_i$ coefficients depend on the number of particles $N$.

One can also investigate the corrections to the vortex modes energies through perturbation theory. Since the two energy levels are perfectly symmetric, only the bias $\Omega$ can lift their degeneracy. At second order, the corrections $\delta E_\pm$ to the $+$ and $-$ modes are given by
\begin{align}
    &\delta E_+^{(2)}= \frac{N(\Delta/2)^2}{-2U(N-1)/3 + 2\hbar\Omega} \\
   \notag &\stackrel{|\hbar\Omega| \ll U}{=}-\frac{3N\Delta^2}{8U(N-1)}-\frac{9N\Delta^2}{8U^2(N-1)^2}\hbar\Omega + \mathcal{O}(\hbar^2\Omega^2)\\
    &\delta E_-^{(2)}= \frac{N(\Delta/2)^2}{-2U(N-1)/3 - 2\hbar\Omega} \\
    \notag&\stackrel{|\hbar\Omega| \ll U}{=}-\frac{3N\Delta^2}{8U(N-1)}+\frac{9N\Delta^2}{8U^2(N-1)^2}\hbar\Omega + \mathcal{O}(\hbar^2\Omega^2).
\end{align}

The first term corresponds to a symmetric energy shift that has no effect on the dynamics. The second term, proportional to $\Omega$, leads to a renormalization of the bias $\pm N_\mathrm{eff}\hbar\Omega$, where
\begin{equation}
    N_\mathrm{eff}= N - \frac{9N\Delta^2}{8U^2(N-1)^2}+\mathcal{O}(\Delta^4/U^4).
\end{equation}

More generally, the perturbative corrections to the diagonal terms can be decomposed into a symmetric contribution (even in $\Omega$), which shifts the global energy, and an antisymmetric contribution (odd in $\Omega$), which renormalizes the effective bias.

\nocite{lerose_2020,dupont_2026,berezin_1975}

\bibliography{bibliography}
\end{document}